\documentclass[
reprint,
 amsmath,amssymb,
 aps,
]{revtex4-2}
\usepackage[T1]{fontenc}
\usepackage[latin9]{inputenc}
\usepackage{amsmath}
\usepackage{graphicx}

\makeatletter

\providecommand{\tabularnewline}{\\}

\makeatother

\usepackage{babel}
\begin{document}
\title{Discontinuous transitions of social distancing in SIR model.}
\author{R. Arazi}
\address{Racah Institute of Physics, The Hebrew University, 9190401 Jerusalem,
Israel}
\author{A. Feigel}
\email{sasha@phys.huji.ac.il}
\address{Racah Institute of Physics, The Hebrew University, 9190401 Jerusalem,
Israel}

\begin{abstract}
To describe the dynamics of social distancing during pandemics, we follow previous efforts to combine basic epidemiology
models (e.g. SIR - Susceptible, Infected, and Recovered) with game and economy theory tools. We present an
extension of the SIR model that predicts a series of discontinuous transitions
in social distancing. Each transition resembles
a phase transition of the second-order (Ginzburg-Landau instability)
and, therefore, potentially a general phenomenon. The first wave of
COVID-19 led to social distancing around
the globe: severe lockdowns to stop the pandemic were followed by a series of
lockdown lifts. Data analysis of the first wave
in Austria, Israel, and Germany corroborates the soundness of the model. Furthermore, this work presents analytical
tools to analyze pandemic waves, which may be extended to calculate
derivatives of giant
components in network percolation transitions and may also be of interest in the context of
crisis formation theories.
\end{abstract}
\maketitle

\section{Introduction}
\label{sec:introduction}

Pandemics are complex medical and socioeconomic phenomena\cite{May1988}: frequent
social interactions benefit both the spread of disease and significant
parts of modern economies\cite{Anderson2020,Atkeson2020}. Rational human behavior during pandemics suggests a balance between
individual efforts to avoid getting infected and the economic costs of protective
measures\cite{Farboodi2020,Alvarez2020}. This balance changes with
time and disease prevalence\cite{Capasso1978}. This work argues that this balance,
together with the corresponding human behavior, may possess
discontinuous transitions similar to the free energy
of a system during the Ginzburg-Landau phase transition and may have some
level of universality\cite{Landau2013,Perc2016,Levy2005,Kelso1984,Encinas2018}. 

Social distancing is an effective tool to mitigate
epidemies\cite{Anderson2020,Atkeson2020}. It consists of self or government-imposed constraints on
interpersonal contacts. Social distancing, however, comes at a significant economic cost in terms of reduced productivity\cite{Diamond1979,Farboodi2020}.

\begin{figure}
\centering{}\includegraphics[scale=0.5]{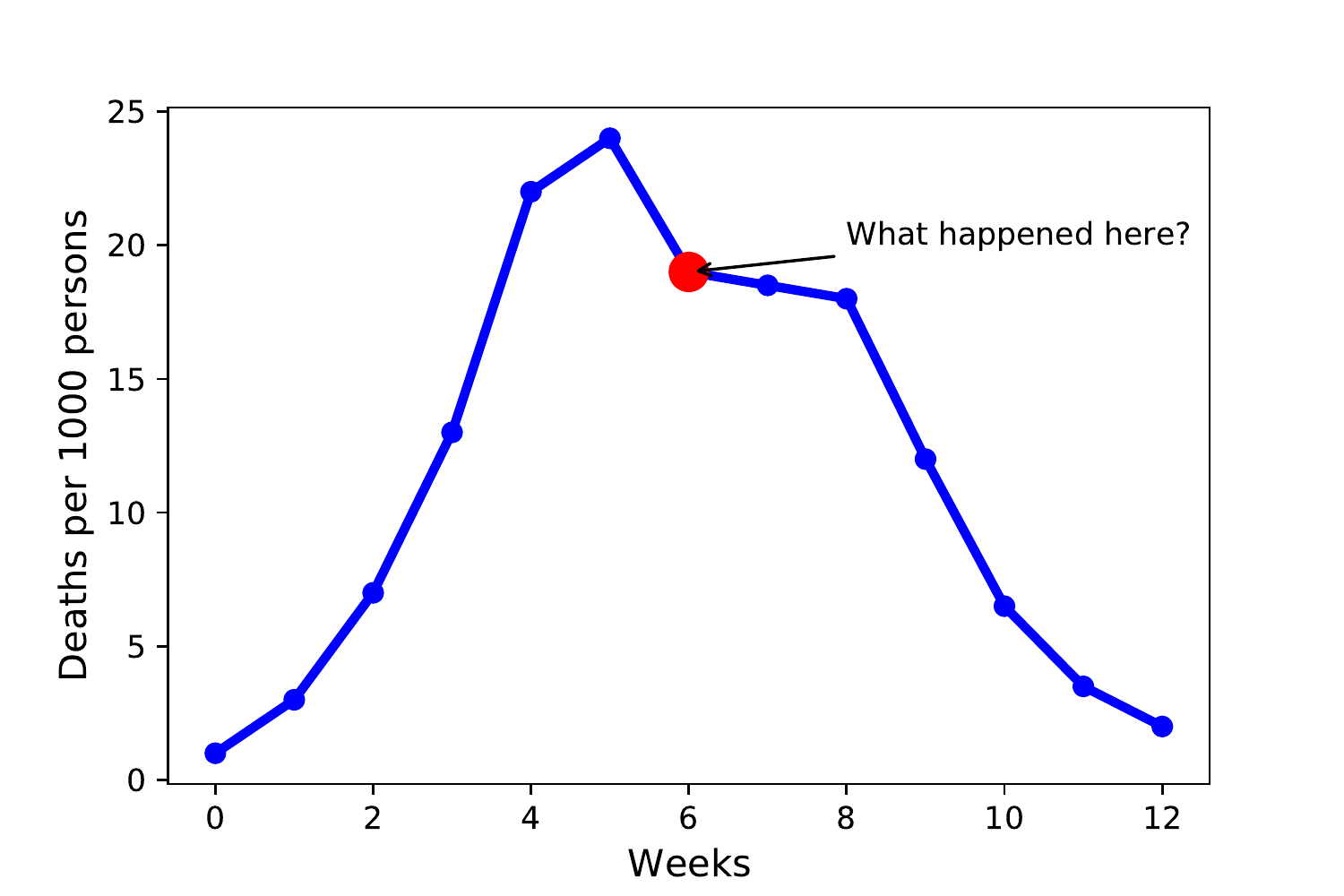}\caption{Mortality
  in England and Wales during the second wave of Spanish flu. The red dot indicates a transition in time
derivative soon after the epidemy peak. This work presents a theory
that describes this type of transition as a rational decision regarding
the optimal value of social distancing in a population. The theory predicts
discontinuous, phase transition-like, changes in social
distancing. We do not claim
that this work provides the only possible explanation of the
discontinuous time derivatives of an
epidemy's dynamics.}
\label{fig:0}
\end{figure}

Social distancing depends on epidemy dynamics and vice
versa\cite{Capasso1978,May1988,Ferguson2007,Bauch2013}. Epidemy
dynamics are known through daily-reported amounts of infected and
deceased persons (see Figure \ref{fig:0} for mortality due to Spanish flu
in England and Wales)\cite{Jordan1927,Chowell2008}. A graph like that
shown in Figure
\ref{fig:0} should reflect changes of social distancing over time.

Reports of confirmed cases or mortalities possess discontinuities in time derivatives. For instance, see the red point in Figure \ref{fig:0} (we will also see similar
phenomena with COVID-19 data). Such transitions may indicate noise in
reported data, changes in testing policy, responses to some extraneous
phenomena, or
a combination of two temporarily and spatially separated epidemy waves. On the other
hand, one can put forward a hypothesis that such transitions indicate abrupt changes
in social distancing practice as a response to reduced levels of epidemy.

For example, consider a hypothesis that discontinuity in time
derivative of reported mortalities (as shown by the red dot in Figure \ref{fig:0})
is a consequence of an abrupt change in social
distancing. The population accepted significant social distancing at the
beginning of the wave but increasingly rejected it after the pandemic passed its
peak. To estimate the time and strength of transition, one needs an
epidemiological model that includes human behavior.

The SIR model\cite{Kermack1927} separates the population into three compartments:
susceptible, infected, and recovered. The flux between these compartments
goes in the order susceptible $\rightarrow$ infected $\rightarrow$
recovered, since susceptible people may become infected during encounters with
 infected individuals. Newly-infected people stay contagious for some time, after which they
stop spreading the disease and become immune (recover) or die. The population is well mixed
and sustains the gas-like interaction of its members.

This work follows many previous efforts to investigate the role of human
behavior during an epidemy\cite{Capasso1978,Ferguson2007,DelValle2005,Fenichel2013a,Fenichel2011,Kremer1996,dOnofrio2009}.
Parameters of SIR\cite{Kermack1927} can depend on disease
prevalence\cite{Capasso1978},  be time
dependent\cite{Kochanczyk2020a}, include spatial effects
\cite{dOnofrio2020,Driessche2000}, be more elaborated by the separation of the susceptible
and infected into sub-compartments\cite{Agaba2017}, include adaptive
mobility\cite{Wang2012}, or include information-related contact patterns\cite{Buonomo2012a,Liu2018,VargasDeLeon2017}. 
Specifically, we have made modifications to the SIR model with economic
tools\cite{Atkeson2020,McAdams2020,Eichenbaum2020,Farboodi2020,Alvarez2020,Toxvaerd2020,Hur2020,Ceddia2013,Morin2013} and
game theory methods\cite{Bhattacharyya2019,Elie2020,Reluga2010}.
Policy\cite{Ferguson2003} and pandemic management is out of the scope of this work.

To proceed, one should associate social distancing with a parameter of
an epidemiological model. Following many previous studies, we will
choose basic reproduction number $R_{0}$ as a measure of both social
distancing and its economic cost\cite{Poletti2009,Reluga2010,Fenichel2011,Wang2015,VenturadaSilva2019,Janssen2018}.

Basic reproduction number $R_{0}$ is the expected number of infected
directly generated by one infected person in a
population where all individuals are susceptible to infection\cite{Diekmann1990}. $R_{0}$
is a measure of social distancing because it is proportional to the frequency of interpersonal
interactions. The economic cost of social distancing, therefore, can be
considered as a function of $R_{0}$\cite{Janssen2018}. Further
,in this work, for the sake of convenience, we will use the inverse of the
basic reproduction number $s_{th}=1/R_{0}$ as the main parameter:
epidemy breaks out only if $R_{0}>0$, thus $s_{th}$ is bounded $0<s_{th}<1$.

In this work,  changes in the value of $s_{th}=1/R_{0}$,
$s_{th}^{(0)}\rightarrow s_{th}^{(1)}$ correspond to the balance between
changes in the final epidemy size $FES$ (the number of new infections from the
current moment till the end of epidemy)\cite{Tanimoto2018,Kuga2019} and changes in economic cost $EC$\cite{Janssen2018}:
\begin{eqnarray}
\label{eq:3}
 FES(s_{th}^{(1)},t)-FES(s_{th}^{(0)},t) = EC(s_{th}^{(1)},t)-EC(s_{th}^{(0)},t).\nonumber\\
\end{eqnarray}
Both $FES$ and $EC$ are defined at time $t$ when individuals (or the
government) make their decision. This time, however, is a function of the subsequent sequence of trajectories of individual decisions.

A major assumption of this work is that decisions regarding transition
$s_{th}^{(0)}\rightarrow s_{th}^{(1)}$ consider only two
possible future trajectories: either $s_{th}^{(0)}$ or
$s_{th}^{(1)}$ and these remain constant until the end of the epidemy. Otherwise,
$FES$ can not be considered entirely as a function of the single value of
$s_{th}$.

The extension of (\ref{eq:3}) in the Taylor series of $\Delta s_{th}$ results
in a non-linear expression similar to the free energy
of a system with Ginzburg-Landau instability.

This work proceeds with the presentation of the SIR model with induced transitions
(SIRIT), an almost analytical treatment of this model, the calibration
of the epidemic and economy parameters of the model (using a time series
of confirmed cases and causalities during the first wave of COVID-19 in
Austria, Israel, and Germany), followed by a discussion of the obtained results
and their implications.

\section{Discontinuous transitions in the SIR model with utility function}
\label{sec:disc-trans-sir}

Here are SIR equations that describe the spread of an epidemy in a population:

\begin{eqnarray*}
\frac{\partial s}{\partial t} & =-\beta is,
\end{eqnarray*}
\begin{eqnarray}
\frac{\partial i}{\partial t} & =\beta is-\gamma i,\label{eq:mainsirit0}
\end{eqnarray}
\begin{eqnarray*}
\frac{\partial r}{\partial t} & =\gamma i
\end{eqnarray*}
where $s(t)$, $i(t)$, and $r(t)$ are the fractions of the population in
the susceptible, infected, and recovered states respectively. The third
equation is redundant since $s+i+r=1$. Rate
$\beta$ includes both the rate of interaction between population members
and the probability of disease transmission during these
interactions. An infected person is contagious during $\gamma^{-1}$ on average.

Both $\beta$ and $\gamma$ may
represent changes in human behavior that
affect the spread of disease. The frequency of social interactions and the level of
self-protection define $\beta$. During severe pandemics like COVID-19, human or government decisions
also affect $\gamma$ by contact tracing and self or government-imposed
quarantine of individuals who are known or suspected to be infected.

If rates $\beta$ and $\gamma$ are constant in time,
eqs. (\ref{eq:mainsirit0}) reduce to:
\begin{eqnarray*}
\frac{\partial s}{\partial t^{*}} & =-si,
\end{eqnarray*}
\begin{eqnarray}
\frac{\partial i}{\partial t^{*}} & =i \left
                                (s-\frac{1}{R_{0}}\right )=i \left
                                (s-s_{th}\right ),\label{eq:mainsirit1}
\end{eqnarray}
where $t^{*}=\beta t$ is dimensionless time and
$R_{0}=\beta/\gamma$ is basic reproduction
number, i.e. the expected number of infections directly generated by one infected person in a
population where all individuals are susceptible to
infection\cite{Diekmann1990}. In addition, $R_{0}$ defines the threshold ratio of susceptible
$s_{th}$ that defines the course of the epidemy: the number of
infected increase if $s>s_{th}$ and decrease if $s<s_{th}$. We will
use $s_{th}$, rather than $R_{0}$, as the main parameter  in
this work.

\begin{figure}
\centering{}\includegraphics[scale=0.4]{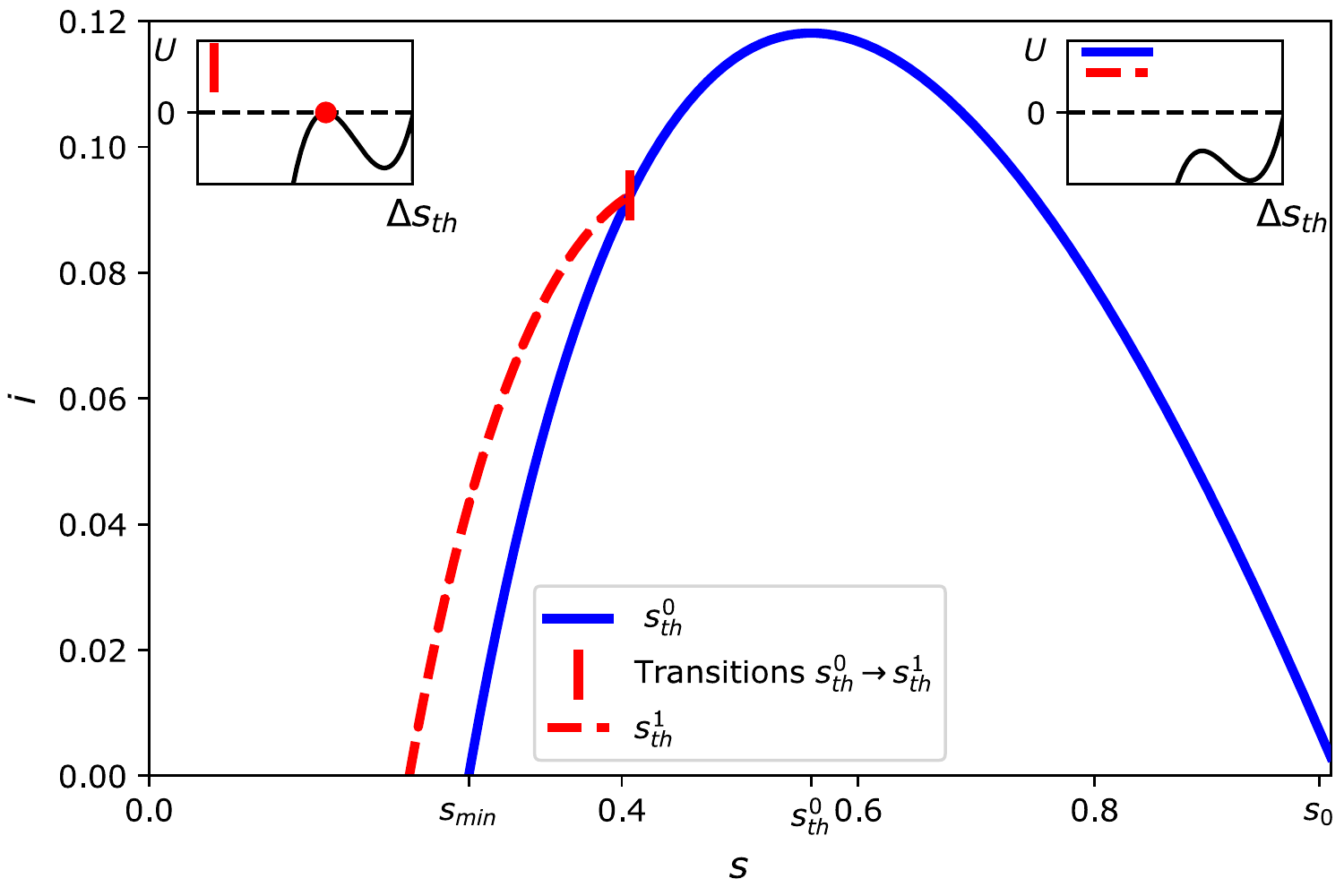}
\caption{ Single transition  $s_{th}^{(0)}\rightarrow s_{th}^{(1)}$. SIR trajectory in $(s,i)$
  space without transition (solid blue). Trajectory starts at
  $(s_{0}\approx 1,i_{0}\approx 0)$. The ratio of susceptible people $s$ in the population
  reduces. The ratio of infected people $i$ increases for $s>s_{th}^{(0)}$ and decreases
  for $s<s_{th}^{(0)}$. Before transition (red bar), utility function $U<0$ for any $s_{th}<s_{th}^{(0)}$. At transition the utility function
  possesses single value $s_{th}^{(1)}$ such that
  $U(s_{th}^{(1)})=0$. Then transition $s_{th}^{(0)}\rightarrow
  s_{th}^{(1)}$ takes place. Immediately after transition again  $U<0$ for any $s_{th}<s_{th}^{(1)}$. The trajectory with transition (dashed
red) converges to $s_{min}^{(1)}$ which is lower than $s_{min}$ of the original
trajectory.}
\label{fig:22}  
\end{figure}

Equation (\ref{eq:mainsirit1}) possesses a solution in $(s,i)$ space, see Figure
\ref{fig:22} and Appendix \ref{sec:analyt-solut-refeq:m}. The form of the trajectory $(s_{t},i_{t})$
depends only on initial values  $s_{th}^{(0)}$ and $(s_{0},i_{0})$. The trajectory starts at $(s_{0},i_{0})$ and advances to
$(s_{min},0)$. The amount of infected people reaches its maximum value at $s=s^{(0)}_{th}$.  

To calculate transition:
\begin{eqnarray}
\label{eq:4}
 s_{th}^{(0)} \rightarrow s_{th}^{(1)},
\end{eqnarray}
following (\ref{eq:3}), we introduce the utility function:
\begin{eqnarray}
\label{eq:5}
U(s_{th},t)=-FES(s_{th},t)+EC(s_{th},t),
\end{eqnarray}
where the final epidemy size $FES$ is:
\begin{equation}
\label{eq:rtr}
FES=\int_{t}^{\infty}idt^{*}.
\end{equation}
and economic cost is some unknown function $EC(s_{th},t)$. The utility
function and its components $FES$ and $EC$ can be considered as
functions of $s$ rather than time $t$ in $(s,i)$ space.

In this work, we consider only the relaxation of social distancing. 
At each moment $t$ the value of $s_{th}^{(0)}$
changes if there exists $s_{th}^{(1)}<s_{th}^{(0)}$ such that:
\begin{equation}
U(s_{th}^{(1)},t)-U(s_{th}^{(0)},t)>0.\label{eq:htrht}
\end{equation}
Both $FES$ and $EC$ are defined at the time $t$ of transition. 

To calculate $s_{th}^{(1)}$ let us expand $U$ into a Taylor
series. Specifically, let us expand  $EC$
to the second order and $FES$ to the third order of $s_{th}$, because $FES$
prevents large changes in $s_{th}$ (a return to the pre-pandemic level of social distancing).

\begin{figure}
\centering{}\includegraphics[scale=0.4]{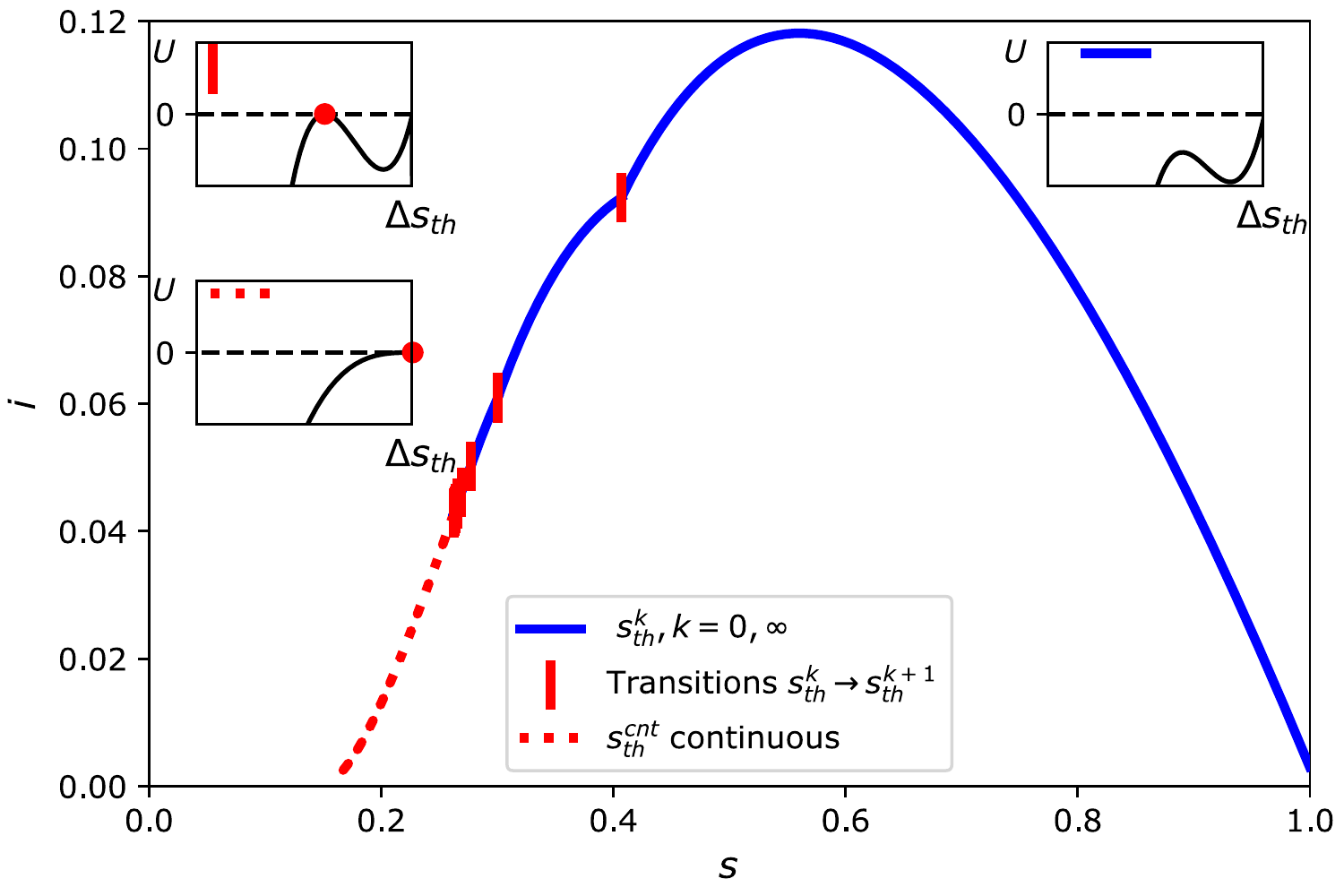}
\caption{Multiple transitions $s_{th}^{(0)}\rightarrow
  s_{th}^{(1)}\rightarrow s_{th}^{(2)}\rightarrow ...\rightarrow
  s_{th}^{(k)}... \rightarrow s_{th}^{\infty}$. Conditions
  $U(s_{th}^{(k+1)})=0$ for $s_{th}^{(k+1)}<s_{th}^{(k)}$ keep hold along
  the trajectory of the SIR model in $(s,i)$ space. This results in a series of
  discontinuous transitions (red bars) that eventually
  converge to some limit values $s_{tr}^{lim}$ and $s_{th}^{lim}$. The time that it takes to pass all
  these infinite number of transitions is finite, as it is any time between
  the two values of $s>s_{min}$.  This results in a Zeno-like phenomenon. The
  final value of $s_{th}^{lim}$ depend on $s_{0},i_{0},s_{th}^{(0)}$ and
  the parameters of economic cost $E^{(1)}$ and $E^{(2)}$. For $s<s_{tr}^{lim}$
  $s_{th}$ changes continuously (dotted red), with
  a utility function that at each moment predicts transition $\Delta s_{th}=0$.}
\label{fig:223c}  
\end{figure}

Economic cost is a general unknown function
$EC(s_{th},t)$. We assume that at each time $t$ it possesses a Taylor expansion:
\begin{equation}
EC(s_{th}+\Delta s_{th},t) = EC(s_{th},t)-E^{(1)}\Delta s_{th}+E^{(2)}\Delta s_{th}^{(2)}.\label{eq:tyy}
\end{equation}
The choice of the signs in (~(\ref{eq:tyy})) assures that $E^{(1)}$ and
$E^{(2)}$ are positive for $\Delta s_{th}<0$. We also assume that parameters
$(E^{(1)}$ and $E^{(2)})$:
\begin{eqnarray}
\label{eq:7}
E^{(1)}=-\frac{\partial EC}{\partial
  s_{th}},E^{(2)}=\frac{1}{2}\frac{\partial^{2} EC}{\partial
  s_{th}^{2}},
\end{eqnarray}
are constant and define the economic cost of changes in $s_{th}$.

The Taylor expansion of final epidemy size~(\ref{eq:rtr}) is:

\begin{eqnarray}
\label{eq:8}
  &&FES(s_{th}+\Delta s_{th},t)=\nonumber \\
  &&FES(s_{th},t)+F^{(1)}\Delta s_{th}+F^{(2)}\Delta
     s_{th}^{2}++F^{(3)}\Delta s_{th}^{3}.\nonumber \\
\end{eqnarray}
where:

\begin{eqnarray}
\label{eq:9}
F^{(1)}=\frac{\partial FES}{\partial
  s_{th}},F^{(2)}=\frac{1}{2}\frac{\partial^{2} FES}{\partial
  s_{th}^{2}},F^{(3)}=\frac{1}{6}\frac{\partial^{3} FES}{\partial
  s_{th}^{2}}.\nonumber \\
\end{eqnarray}
$F^{(1)}$, $F^{(2)}$ and $F^{(3)}$ depend on position along the
trajectory $(s_{t},i_{t})$.

Coefficients $F^{(1)}$, $F^{(2)}$ and $F^{(3)}$ can be derived as analytical
functions of $s_{t}$ and $s_{th}^{(0)}$. Following (\ref{eq:mainsirit1}) and (\ref{eq:rtr}), $FES$ at any time $t$ is:
\begin{equation}
FES=\log s_{min}- \log s_{t},
\end{equation}
Thus $F^{(i)}$ are derivatives of $s_{min}$ due to $s_{th}$.
$s_{min}$ can be expressed using the Lambert W function\cite{Lehtonen2016,Wang2010,Reluga2004,Britton2010}:
\begin{equation}
s_{min}=-s_{th}^{(0)}W\left(-\frac{s_{0}}{s_{th}^{(0)}}\exp\left[-\frac{i_{0}+s_{0}}{s_{th}^{(0)}}\right]\right).\label{eq:fsffsf}
\end{equation}
Derivatives of the Lambert W function can be calculated
analytically. $s_{min}$ does not change along the trajectory, thus at any time before
transition one can change $(s_{0},i_{0})\rightarrow
(i_{t},s_{t})$. Derivatives of  $s_{min}$ due to $s_{th}$ depend on
$(s_{t},i_{t})$. Finally, $F^{(1)}$, $F^{(2)}$, and
$F^{(3)}$ are functions of $s_{t}$, $s_{th}$, (see Appendix \ref{sec:deriv-lamb-w}).

\begin{figure}
\centering{}\includegraphics[scale=0.5]{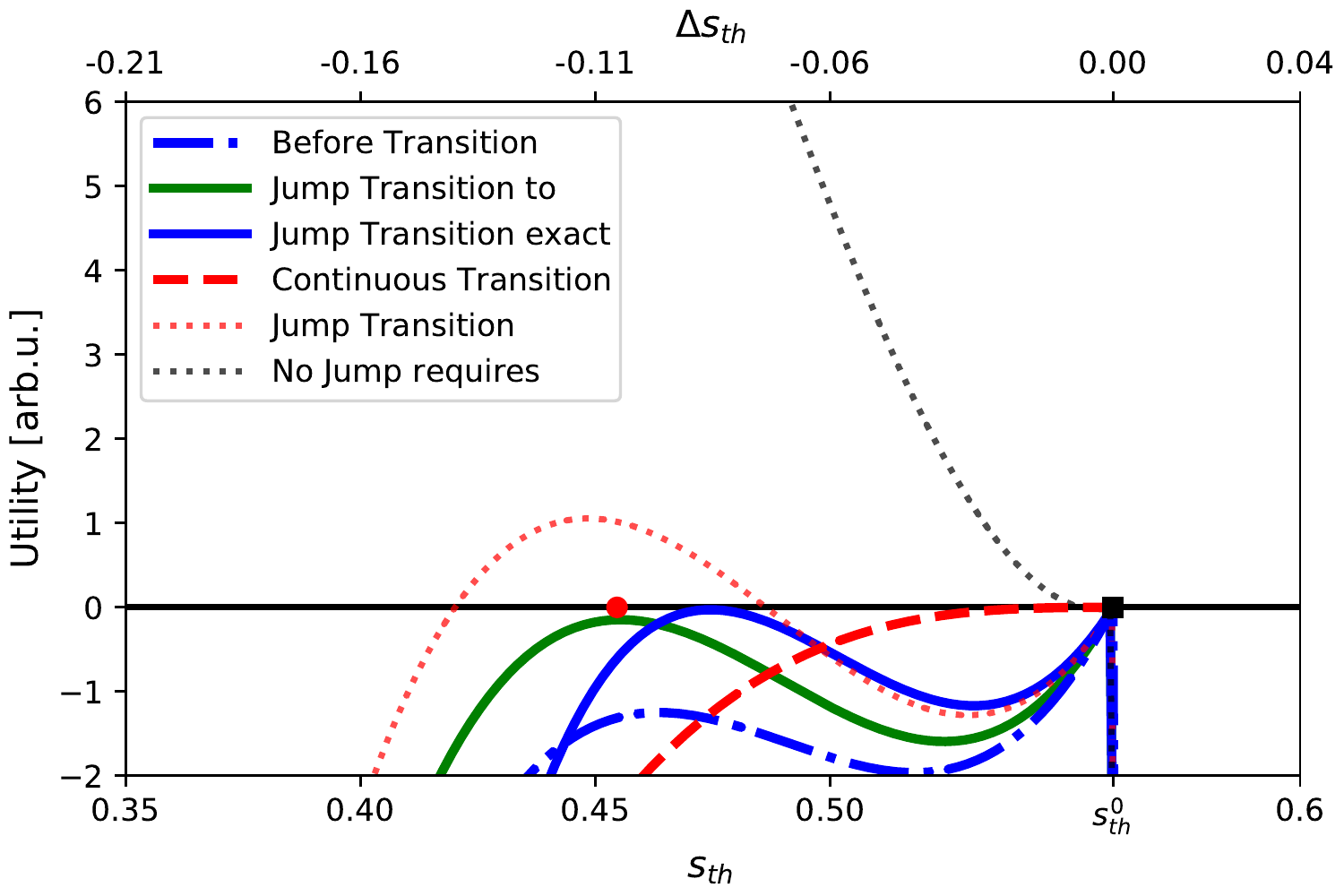}\caption{Utility
  as a function of $s_{th}$ before, during, and after
transition takes place. A measure
of social distancing is $0<s_{th}<1$. Transition occurs between current
value of $s_{th}^{(0)}$ (black
square) to its new value $s_{th}^{(1)}$ (red circle) when $U(s_{th}^{(1)})>U(s_{th}^{(0)})$
($U(s_{th}^{(0)})=0$). Before transition $U(s_{th}^{(0)})$
is the highest value of the utility function (dashed-dotted blue). Then there are two possibilities:
discontinuous change $s_{th}^{(1)}\protect\neq s_{th}^{(0)}$ (solid
green line) or continuous change $s_{th}^{(1)}\approx s_{th}^{(0)}$
(dashed red line). Two cases (dotted lines) that make possible the change
of $s_{th}$ to many values do not exist because either continuous
or discontinuous transition occurred before. This work considers only
a reduction in the level of social distancing, which corresponds to a reduction of $s_{th}$.
Utility function is approximately a cubic function of $\Delta s_{th}$
(solid green). The transition predicted by the exact calculation (solid
blue)  for this work predict insignificant changes in the time and strength
of the transition.}
\label{fig:1}
\end{figure}

The main proposition of this work is that utility function~(\ref{eq:5})
possesses Ginzburg-Landau-like instability, see Figure \ref{fig:1}. The utility function (\ref{eq:5}), taking into account (\ref{eq:tyy}) and
(\ref{eq:8}), is $U(s_{th}+\Delta s_{th})=U(s_{th})+\Delta U$ where:
\begin{equation}
\Delta U=-(F^{(1)}+E^{(1)})\Delta s_{th}-(F^{(2}-E^{(2)})\Delta s_{th}^{2}-F^{(3)}\Delta s_{th}^{3}.\label{eq:rgtgt}
\end{equation}
is the third-degree polynomial of $\Delta s_{th}$. First, no
transition occurs if $\Delta U<0$ for all $\Delta s_{th}<0$. Second, discontinuous
change in $s_{th}^{(0)}\rightarrow s_{th}^{(1)}$ takes place if there is single value $\Delta
U(s_{th})=0$ for all $\Delta s_{th}<0$.
Third,  $s_{th}$ changes continuously when derivatives
of (\ref{eq:rgtgt}) vanish near $\Delta s_{th}=0$. To prevent
transitions with $\Delta s_{th}>0$ we assume that the economic cost of additional
 social distancing is high and overrides the potential reduction
of $FES$.

Discontinuous transition occurs when there exists a single $\Delta
s_{th}<0$ root for $\Delta U=0$ (\ref{eq:rgtgt}). This condition requires the determinant of quadratic function $U/\Delta
s_{th}$ (\ref{eq:rgtgt}) to vanish:
\begin{equation}
4F^{(3)}(F^{(1)}+E^{(1)})=(F^{(2)}-E^{(2)})^{2}.\label{eq:25253}
\end{equation}
This is the fourth-order polynomial of $\log s_{t}$ because  $F^{(1)}$, $F^{(2)}$, and
$F^{(3)}$ are polynomials of $\log s_{t}$ of the first, second, and 
third degrees correspondingly, see Appendix \ref{sec:final-expressions-b}.

Following (\ref{eq:rgtgt}), transition strength $\Delta s_{th}$ is:
\begin{equation}
\Delta s_{th}=\frac{F^{(2)}-E^{(2)}}{2F^{(3)}}.\label{eq:gdfgdf}
\end{equation}
Then:
\begin{eqnarray}
s_{th}^{(1)} =s_{th}^{(0)}+\Delta s_{th},
\end{eqnarray}
is the new value of $s_{th}$.

To calculate when transition takes place on a trajectory
$(s_{th}^{(0)},s_{0},i_{0})$ one should solve the 4th order
polynomial (\ref{eq:25253}) for $s_{tr}$ and take the closest to $0$ negative root. 
One can use (\ref{eq:gdfgdf}) to calculate new value
$s_{th}^{(1)}$ and continue the trajectory $(s_{th}^{(1)},s_{tr},i(s_{tr}))$.

Previous results can be presented as two functions:
\begin{eqnarray}
  \label{eq:fgd}
  s^{(1)}_{tr}&=&T_{1}(s_{0},i_{0},s_{th}^{(0)},E^{(1)},E^{(2)}),\nonumber \\
  s_{th}^{(1)}&=&T_{2}(s_{0},i_{0},s_{th}^{(0)},E^{(1)},E^{(2)}).
\end{eqnarray}
 $T_{1}$ defines where transition takes place $s$=$s_{tr}^{(1)}$ while
 $T_{2}$ defines the transition strength $s_{th}^{(0)}\rightarrow s_{th}^{(1)}$.

Consider the single transition in Figure \ref{fig:22}. This trajectory consists
of the points $(s_{t},i_{t})$, starts
at $(s_{0},i_{0})$ and proceeds to lower values of $s$. At some moment
condition (\ref{eq:25253}) alarms and a new value of $s_{th}$ appears.

For a single transition, one can consider an inverse problem - to
calculate  $E^{(1)}$ and $E^{(2)}$ if the transition position
$s_{tr}$  and its strength $\Delta s_{th}$ are known.

The trajectory $(s_{t},i_{t})$ may include multiple transitions.
Condition (\ref{eq:25253}) may occur after the first transition and so
on. To calculate these transition one should repeatedly apply
(\ref{eq:fgd}) with change $(s_{th}^{(k)},s_{k},i_{k})\rightarrow (s_{th}^{(k+1)},s_{k+1},i_{k+1})$. 

Many transitions (\ref{eq:fgd}) result in a Zeno-like phenomenon - the
system makes an infinite number of transitions in a finite amount
of time, see Figure \ref{fig:2}. Repeated application of $T_{1}$ and
$T_{2}$  (\ref{eq:fgd}) results
in convergence of $s_{th}$ and $s_{tr}$ to some limit values  ($s_{th}^{lim}$ and $s_{lim}$
respectively). 
The time to reach $s_{th}^{lim}$ is finite, as any time
between the two values of $s>s_{min}$, see (\ref{eq:ertre}).

For $s<s_{tr}^{lim}$, the utility function preserves the continuous
transition state.  At $s_{th}^{lim}$ (\ref{eq:fgd})
predicts $\Delta s_{th}=0$, see Figure \ref{fig:1}. At the region of continuous transitions, at each moment
equation:
\begin{equation}
F^{(1)}=-E^{(1)}\label{eq:fgfdf}
\end{equation}
defines $s_{th}$, and (\ref{eq:mainsirit1}) is solved numerically.
Otherwise, if $s_{th}$ remains constant the utility
function would make possible many values of $s_{th}$ with $U(s_{th})>0$,
see Figure \ref{fig:1}.

The single transition requires an assumption that the economic cost is
a function of the basic reproduction number. For multiple transitions, we assume that economic cost is the same
for all transitions. To check the validity of these and other assumption of
the model we proceed to fit the first wave of COVID-19.

\section{Fit}

The main purpose of this section is to show that the model of
discontinuous transitions of social distancing (see section \ref{sec:introduction})
may fit the first COVID-19 wave data. Full
optimization of the COVID-19 data fit and its validation is out of
the scope of this work. To proceed we want to make our model
compatible with the COVID-19 data.

During the COVID-19 pandemic, there are daily worldwide reports\cite{Dong2020}
of confirmed cases and  deaths. Confirmed cases are
detected infections, which are a fraction of the total number of infected people. The ratio between confirmed and total
coronavirus cases is unknown and may vary from country to country or
even change in time due to changes in test policies. Nevertheless, some
countries may be similar to each other\cite{Mossong2008}.

For the sake of the fit, first, we should introduce new parameters that represent unknowns
of the data. Second, we should consider dynamics in time-space rather
than $(s,i)$ space.

Let us rewrite (\ref{eq:mainsirit0}) as:
\begin{eqnarray*}
\frac{\partial s}{\partial t} & =-\beta AsI,\quad A=\frac{A'}{N},
\end{eqnarray*}
\begin{eqnarray}
\frac{\partial I}{\partial t} & =\beta I(s-s_{th}),\label{eq:mainsirit2}
\end{eqnarray}
\begin{eqnarray*}
\frac{\partial D}{\partial t} & =\beta s_{th}NAIM
\end{eqnarray*}
where $s$ remains to be fraction of susceptible, $I$ reported confirmed cases and $D$ is reported deaths due
to the epidemy. $A'$
is a ratio between actual and reported confirmed cases and $N$ is the population
size. Thus $D=MNr$, where $M$ is infected fatality rate (IFR). When population
size $N$ and $A'$ remain constant it is convenient to unite them
into a single parameter $A=\frac{A^{'}}{N}$. Eqs.
(\ref{eq:mainsirit2}) converge to (\ref{eq:mainsirit0}) if $A'=1$, with 
the only difference being that the third equation addresses deceased instead of recovered. Parameter $\beta$ and $s_{th}$ are the same as in (\ref{eq:mainsirit0}).

To proceed with a fit we assume that only those parameters that are related
to human behavior can change with time. Thus $\beta$ and $s_{th}$ change
with time, while $A'$, $N$ and $M$ remain constant (these parameters
relate to testing policies and disease clinics.). This choice fits our
purpose to
show that COVID-19 data can be explained by changes in social
distancing. In reality,
however, all parameters of the SIR model may be continuous functions of
time or the state of the epidemy. The implications of this assumption are addressed in
the discussion section \ref{sec:discussion}.

\begin{figure}
\centering{}\includegraphics[scale=0.5]{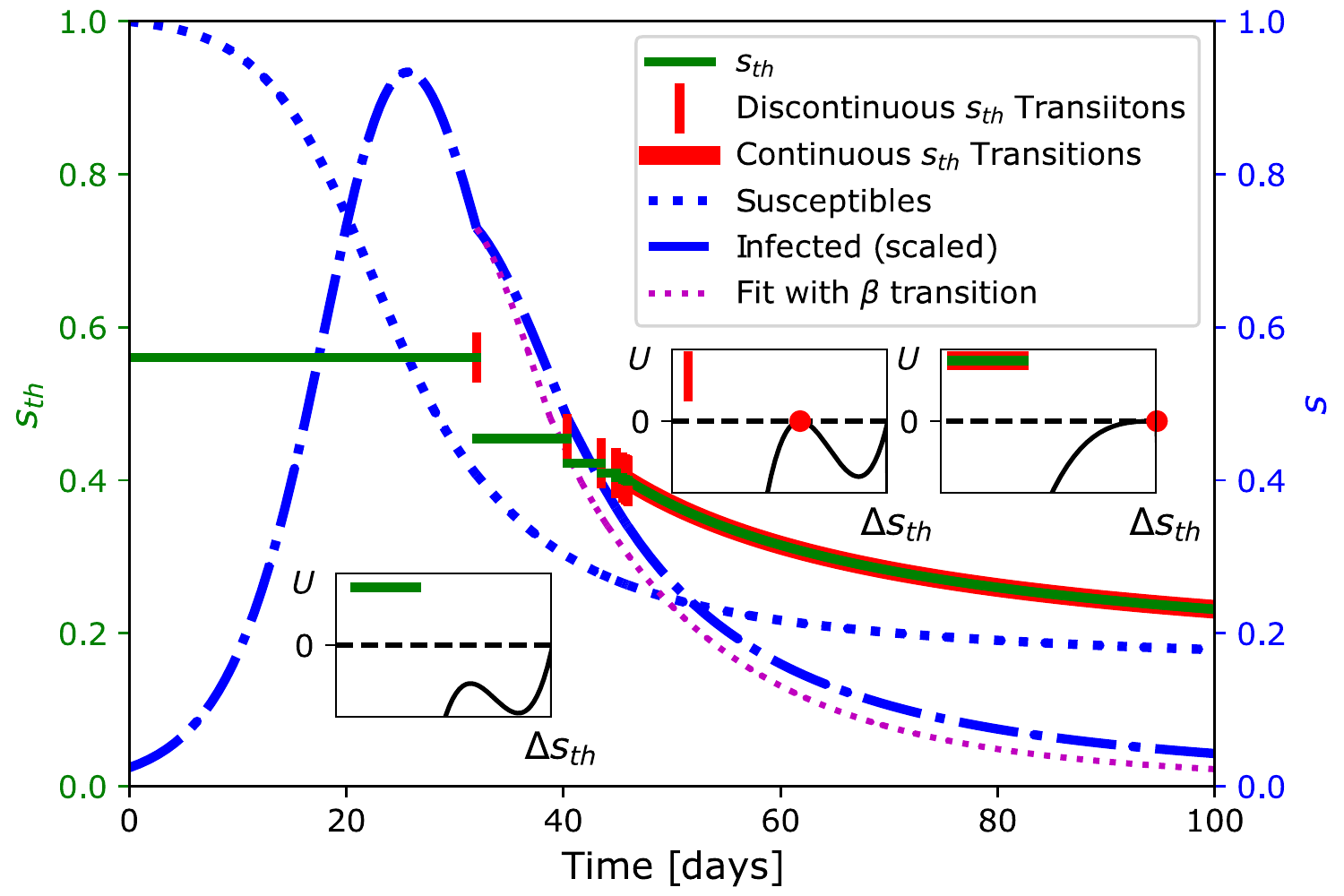}\caption{Epidemy wave dynamics with social distancing transitions. SIR predicts
wave-like behavior of infected individuals (dash-dotted blue) accompanied
by a reduction in the number of susceptible people (dotted blue). A utility function
causes changes in social distancing parameter $s_{th}$ (solid green).
Transition (red bars) occurs when there exists a new value of $s_{th}^{(1)}$
such that the utility function $U(s_{th}^{(1)})>0$. The forms of the utility function
before and during transition are shown in the subplots. A region of continuous
transitions follows a series of discontinuous ones. Changes in $s_{th}$ cause
time derivative discontinuities in the dynamics of the infected and
susceptible numbers.  These are especially prominent in the first transition (outermost left
red bar) soon after
the epidemy's peak. In time space population trajectory depends on whether
the change of $s_{th}=\gamma/\beta$ occured by the change of $\gamma$
(dashed-dotted blue) or $\beta$ (dotted magenta). The difference is
insignificant for the purpose of this work.}
\label{fig:2}
\end{figure}

See Figure \ref{fig:2} for trajectory in time, corresponding to
trajectory in  $(s,i)$ space in Figure \ref{fig:223c}. The trajectory in
time space depends on $\beta$ and $s_{th}$, rather than only $s_{th}$. Thus
transition $s_{th}^{(k)}\rightarrow s_{th}^{(k+1)}$ may be interpreted as
a change in $\beta$, in $\gamma$ or in both $\beta$ and
$\gamma$. Choosing every transition as a change in $\gamma$ (dash-dotted blue) or change
in $\beta$ (dotted magenta) results in similar trajectories, with
differences that are below
resolution of this work. It is true that if the first transition occurs
soon after the maximum number of infected is reached - other conditions may
cause greater differences between $\beta$ and $\gamma$ fits. 

The fit proceeds in the following steps: first, a small region around the greatest number of
infected, see Figure \ref{fig:3}, is used to calibrate $s_{0},I_{0},\beta,A$.
The initial value $s_{th}^{(0)}=\gamma/\beta$. During the first fit, $\gamma=0.26$ is constant. This choice can be quite arbitrary in
the boundaries of reported COVID-19 $1/7<\gamma\;[day^{-1}]<1$
values\cite{Boehmer2020}. We remind you that it also can be affected by
contact tracing and isolation policies. 

Second, one detects a discontinuous change
in the time derivative of reported confirmed cases, see
the outermost left red bar in Figure \ref{fig:3}. It defines $t$ and
$s_{tr}$ of the first transition. New values of $\beta$ and $\gamma$
were fitted for a confirmed case during the period of about 20 days after the
transition. In all cases, the fit predicted $\beta$ to remained
unchanged while $\gamma$ becomes a new value. Thus $s_{th}^{(0)}$ becomes $s_{th}^{(1)}$. Third,
eq. (\ref{eq:fgd}) is solved for $E^{(1)}$ and $E^{(2)}$ (e.g.
using the Nelder-Mead method), when $s_{tr}^{(1)}$ and $s_{th}^{(1)}$ are
the values from the previous step. Fourth, next transitions
$s_{tr}^{(2)}$ and $s_{th}^{(2)}$,
 are calculated using (\ref{eq:fgd}) and (\ref{eq:fgfdf}). This
 procedure is repeated to calculate discontinuous transitions
 $s_{tr}^{(k)}$ and $s_{th}^{(k)}$ till the limit $s_{tr}^{lim}$ is reached. The
 continuous changes of $s_{th}$ are calculated using (\ref{eq:fgfdf}).

To fit casualties, an effective population size $N$ is chosen to fit
reported coronavirus deaths on the 100th day of the first wave. It
predicted a quite small
$N$ less than $1/10$ of the Austrian population.  IFR is an average probability of an
infected person to die (for instance, reported COVID-19 IFR
in Germany $M\approx 0.37\%$\cite{Streeck2020}). This result will be addressed during the discussion below. Besides, there exists
some time shift between calculated and reported coronavirus
deaths.

The complete dynamics of the first wave confirmed cases and
susceptible people, see Figure \ref{fig:3}, follow eqs. (\ref{eq:mainsirit2}) and
the fitted $s_{0},I_{0},s_{th},\beta,A,N$ together with the transitions'
locations $t_{i},s_{tr}^{(i)}$ and strengths $s_{th}^{i}$. See Figure
\ref{fig:3} for changes in $s_{th}.$

Two alternative fits of Austrian COVID-19 data are shown in Figure \ref{fig:4}.
The first demonstrates the sensitivity of the fit to the choice of the
first transition. The second demonstrates that for any first transition
the entire curve can be fitted if $E^{(1)},E^{(2)}$are fitted separately.

The results for Germany and Israel are summarized in Figures \ref{fig:5}
and \ref{fig:6} together with Table \ref{tab:1}, which summarizes the results for all three countries. All
countries demonstrated a low size of the effective population. One of the
assumptions was that $s_{0}\approx1$. In the case of Israel, it was required to
be constrained during the fit. Neither of the deviations from the fit refutes the main results
of this work.

\begin{widetext}
\onecolumngrid
\begin{figure}
\centering{}\includegraphics[scale=0.4]{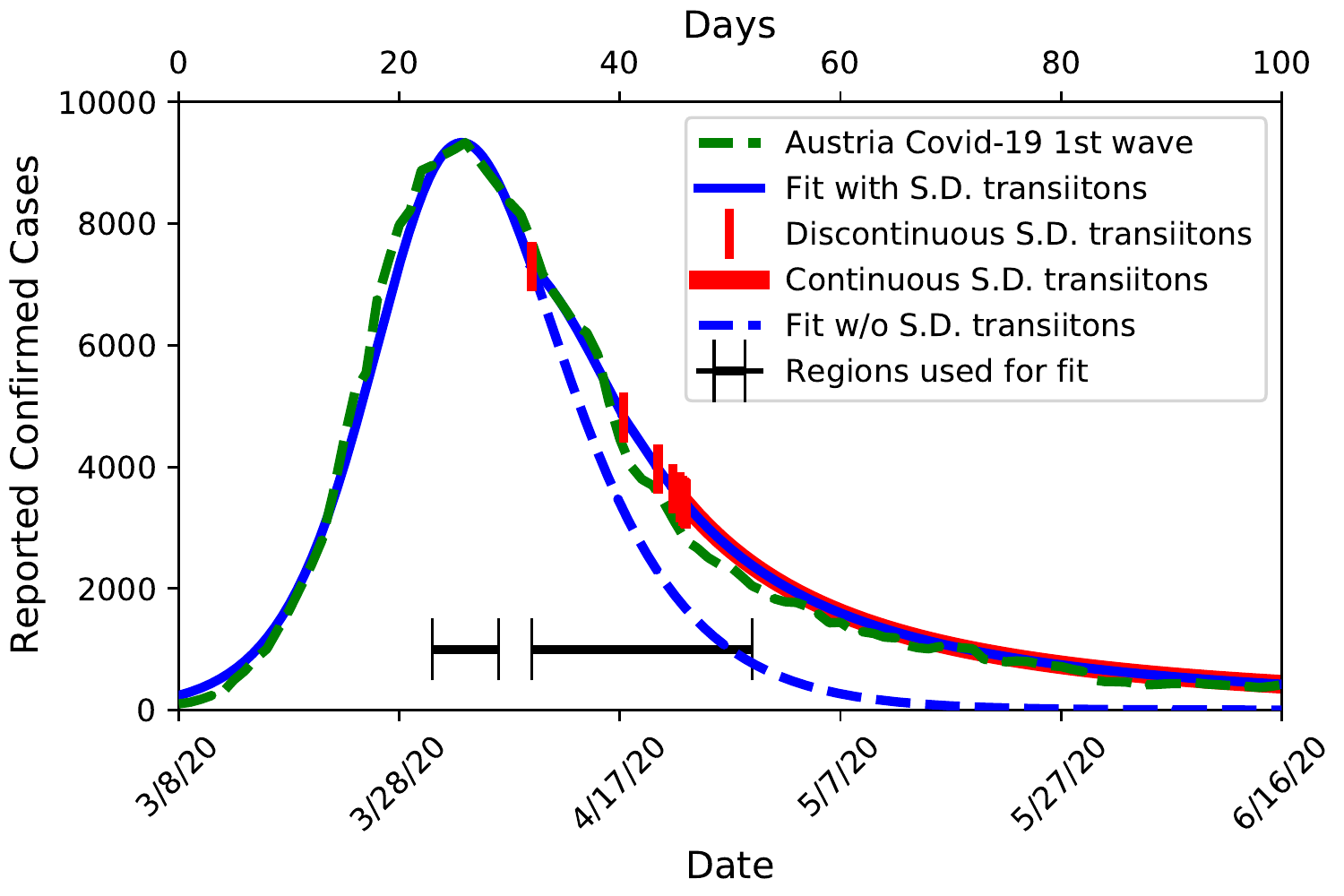}\includegraphics[scale=0.4]{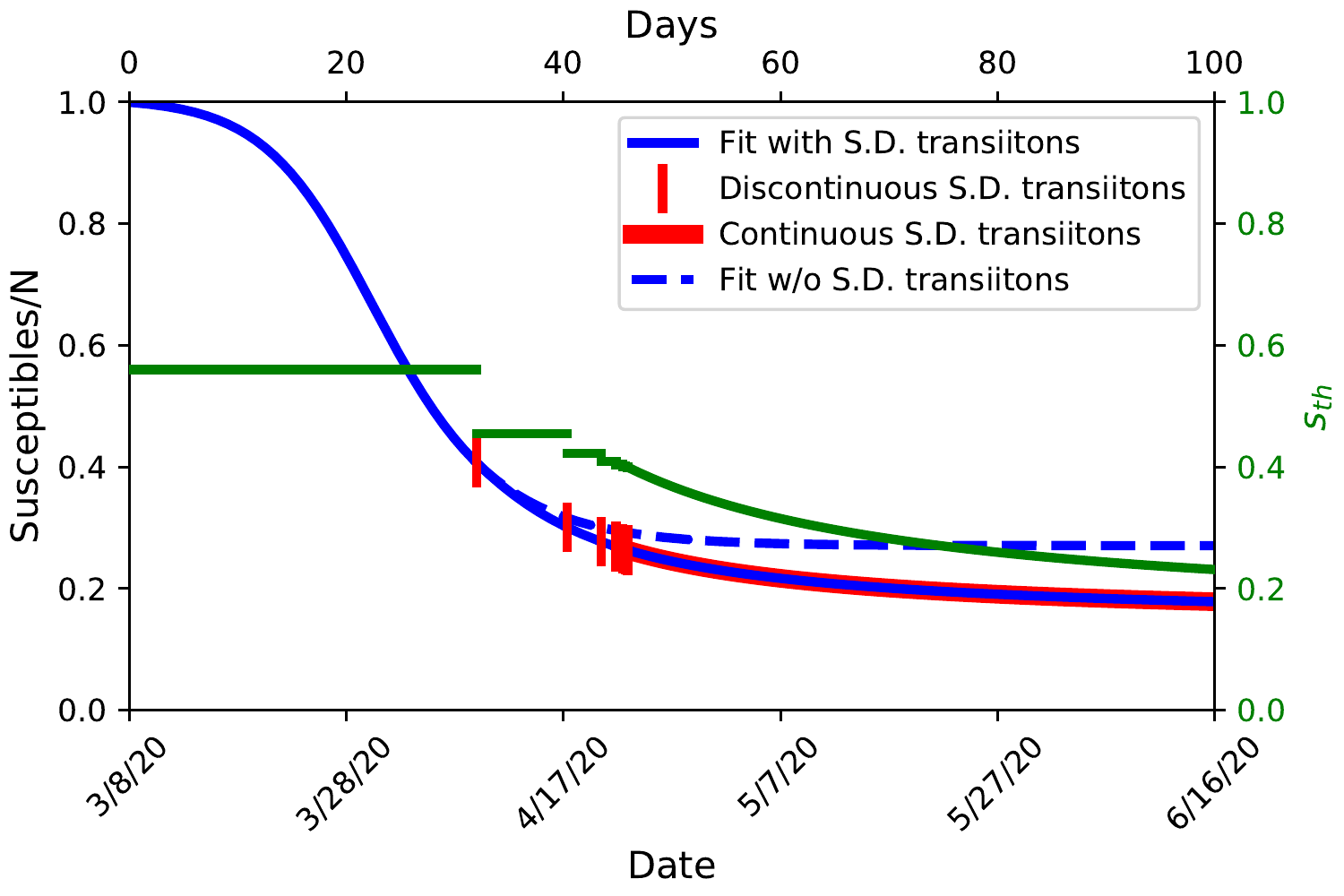}\includegraphics[scale=0.4]{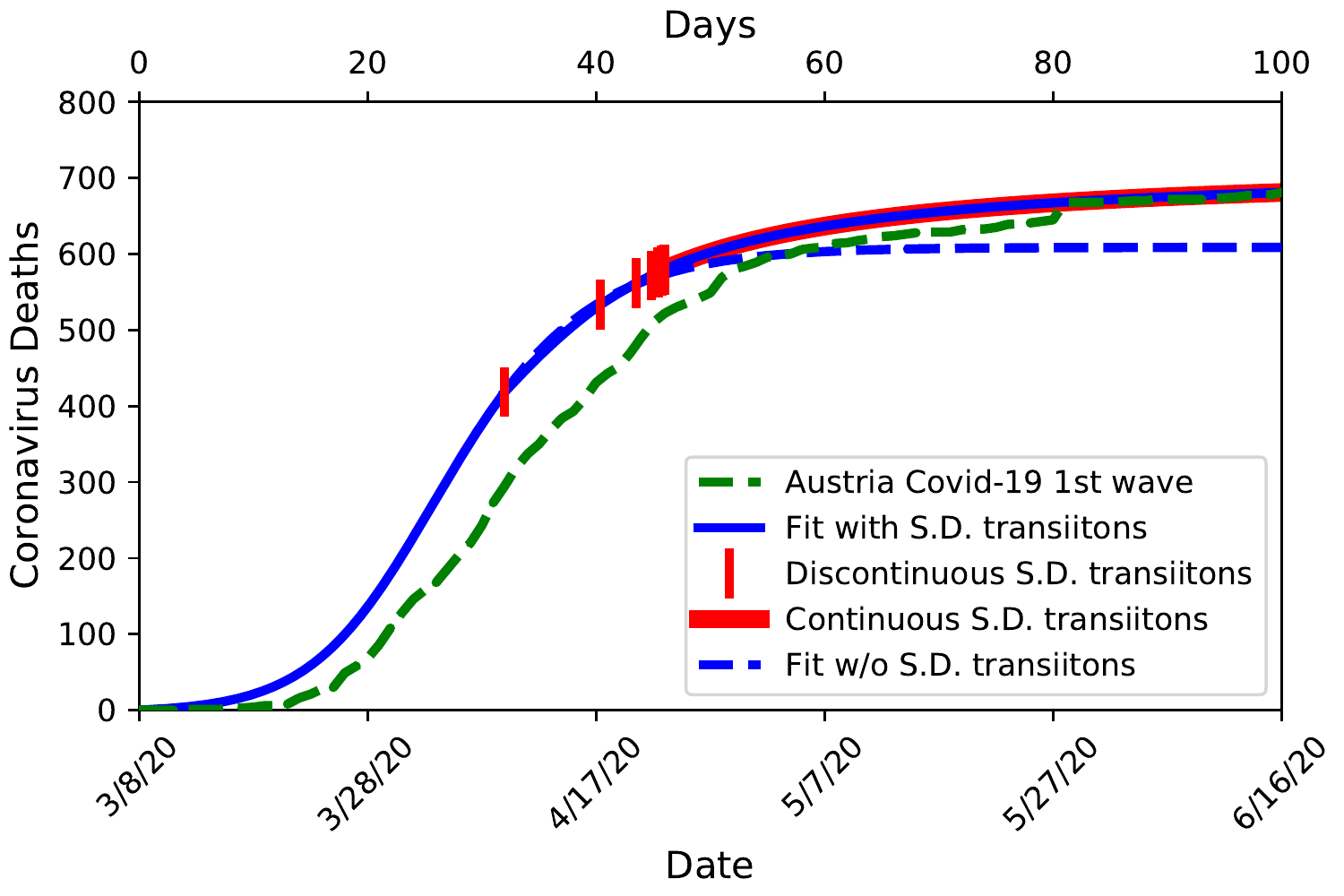}
\caption{Fit of Austria COVID-19 first wave with multiple social distancing transitions.
The purpose of the fit is to show that there is a possibility
to fit the real data using many transitions. A) Confirmed cases -
reported (dashed green) and calculated with many transitions (solid
blue). Before transitions (red bars) classical SIR fits well with the reported
confirmed cases. There is a significant deviation of SIR from reported
cases after the first transition (dashed blue). The SIRIT model with many
transitions fits well with the entire range of the first wave, though the fit
was obtained using a small range around the peak of confirmed cases
and characteristics of the first transition (horizontal error bars).
B) Susceptible and social distance parameter $s_{th}$. At each transition
$s_{th}$ changes its value. A series of discontinuous transitions is
followed by a continuous change region. C) Coronavirus deaths. There
exists a time delay between reported and calculated deaths. This delay
can be explained by the long course of COVID-19.}
\label{fig:3}  
\end{figure}
\twocolumngrid
\end{widetext}

Analysis of the first wave predicts a small, less than $1/10$, effective
population
size in all three countries studied, see Table \ref{tab:1}. An estimate of
effective population size depends on the choice of $M$ - infected
fatality rate (IFR). Increase/reduction in $M$ causes a proportional
reduction/increase in effective population size 
$N$ and predicted ration $A'$ between reported and real numbers of
infected. These values in Table \ref{tab:1} correspond to
$M=0.1\%$.  The reported value of $M$ for Germany is
$0.37\%$\cite{Streeck2020}. Thus $N$ and $A'$ maybe about $\times 4$
lower than in Table \ref{tab:1}. Nevertheless $M$ as low as $0.17\%$
were reported\cite{Bendavid2020}. All other predictions or results of
this work, including the graphs, are independent of $M$. Mortality
rate $M>0.3\%$ causes non-physical $A'<1$ in the case of
Israel. 

The first waves of COVID-19 in Austria, Germany, and Israel were fitted
using SIRIT in two different ways. The first was a series of discontinuous
transitions with constant economy parameters. The parameters $E^{(1)},E^{(2)}$
were fitted by the first transition. In the second, economic weights were
fitted for every candidate transition (deviation from SIR
model). Both these scenarios include discontinuous changes in social
distancing and possess the same first transition.

\section{Discussion}
\label{sec:discussion}

This work describes Ginzburg-Landau-like
instability in the SIR epidemiological model extended with
time-dependent human behavior. The utility function models rational
decision making regarding the optimal level of social distancing. First,
we describe the discontinuous dynamics of this model. Second, we try to show that reported infections or deaths
during the COVID-19 pandemic may
include the evidence of a predicted discontinuous transition. The first task
allows rigorous treatment. The second task requires many assumptions that we will discuss shortly.

Let us start with a discussion of the choice of the utility function and
the corresponding decision-making process.  The utility function consists of
two parts - the cost of the epidemy and the economic cost of social
distancing.

Following several previous works, we take final epidemy size $FES$ as the
main cost of the epidemy. $FES$ according to the SIR model in $(s,i)$ space, in
addition to the current state of the epidemy $(s_{t},i_{t})$, depend on the single
parameter - basic reproduction number $R_{0}$. The price of pandemics,
however, can go beyond the final epidemy size $FES$.
For instance, one may include time derivatives of the number of infected people as
a psychological factor that affects individual decision making. There
is a lot of room to make the utility function more complicated, but this
work demonstrates that discontinuous dynamics may be achieved even with a basic model.

The economic cost is assumed to be a function of $s_{th}=1/R_{0}$. Basic
reproduction number $1/s_{th}=R_{0}$ also  serve as a social distancing
parameter. This choice does not change the major predictions or analytical
developments of this work. In our opinion, the parameter $0<s_{th}<1$ that can be
compared with the fraction of susceptible people in a population serves
better the purpose of this work.

The main purpose of a utility function is to represent decision
making. This work assumes that decision making considers every
transition $s_{th}^{(k)}\rightarrow s_{th}^{(k+1)}$ as a single one,
suggesting that the new value $s_{th}^{(k+1)}$ preserves till the end of
epidemy. This is a major assumption of this work because  elaborate
decision making should consider future possible changes in
$s_{th}^{(k+1)}$.

The specific choice of the utility function makes the possible rigorous treatment
of a single transition of social distancing. Multiple transitions
require  additional assumptions around economic costs which change from transition
to transition. Economic cost is assumed to be constant with time during the fit of COVID-19 data.

The ratio between total and confirmed infected  $A'$, population size $N$,
and infected fatality rate (IFR) $M$ is assumed to be constant during the
first wave of COVID-19. It is a very strong assumption, especially
regarding ratio $A'$, which depends on the amount of the
tests. Nevertheless, one can hope that $A'$ preserves for some time while test policies are the same. Especially important
for this work is the time from the maximum number of infected till the
first predicted transition, see Figure \ref{fig:3}.

All these assumptions provided a successful fit of COVID-19 data in
Austria, Germany, and
Israel. Austria and Israel are countries with similar population
sizes and with similar policies during the initial stages of the first wave.
Germany is a country with a population about $\times10$ the size of
the other two countries, which still
demonstrated SIR-like behavior during the first wave. 

All three  countries studied entered a
transition soon after the epidemy started to decline. During the fit, this work can not distinguish
between the transition of social distancing $s_{th}$ and changes in
parameters that were assumed to be constant, for instance, $A'$. Thus
,transition in social distancing remains a hypothesis.

An interesting conclusion of the fit is that the first transition of
social distancing $s_{th}^{(0)}\rightarrow s_{th}^{1}$ corresponds to
the change of $\gamma$ rather than $\beta$.  $\beta$ remains constant
during the first transition of $s_{th}=\gamma/\beta$.  Generally, human behavior is associated with changes
in $\beta$ that are directly connected to the frequency of interactions
and behavior during such interactions. Nevertheless, during extreme
events such as COVID-19, the contagious period $\gamma^{-1}$ can be shortened due to contact tracing and the isolation of the confirmed
or possibly infected. Some studies highlight
the importance of $\gamma$ changes\cite{Gans2020}. It is important to state that the general
theory of social distancing transitions in section
\ref{sec:disc-trans-sir} is not affected by this finding.

\begin{widetext}
\onecolumngrid

\begin{figure}
\centering{}\includegraphics[scale=0.5]{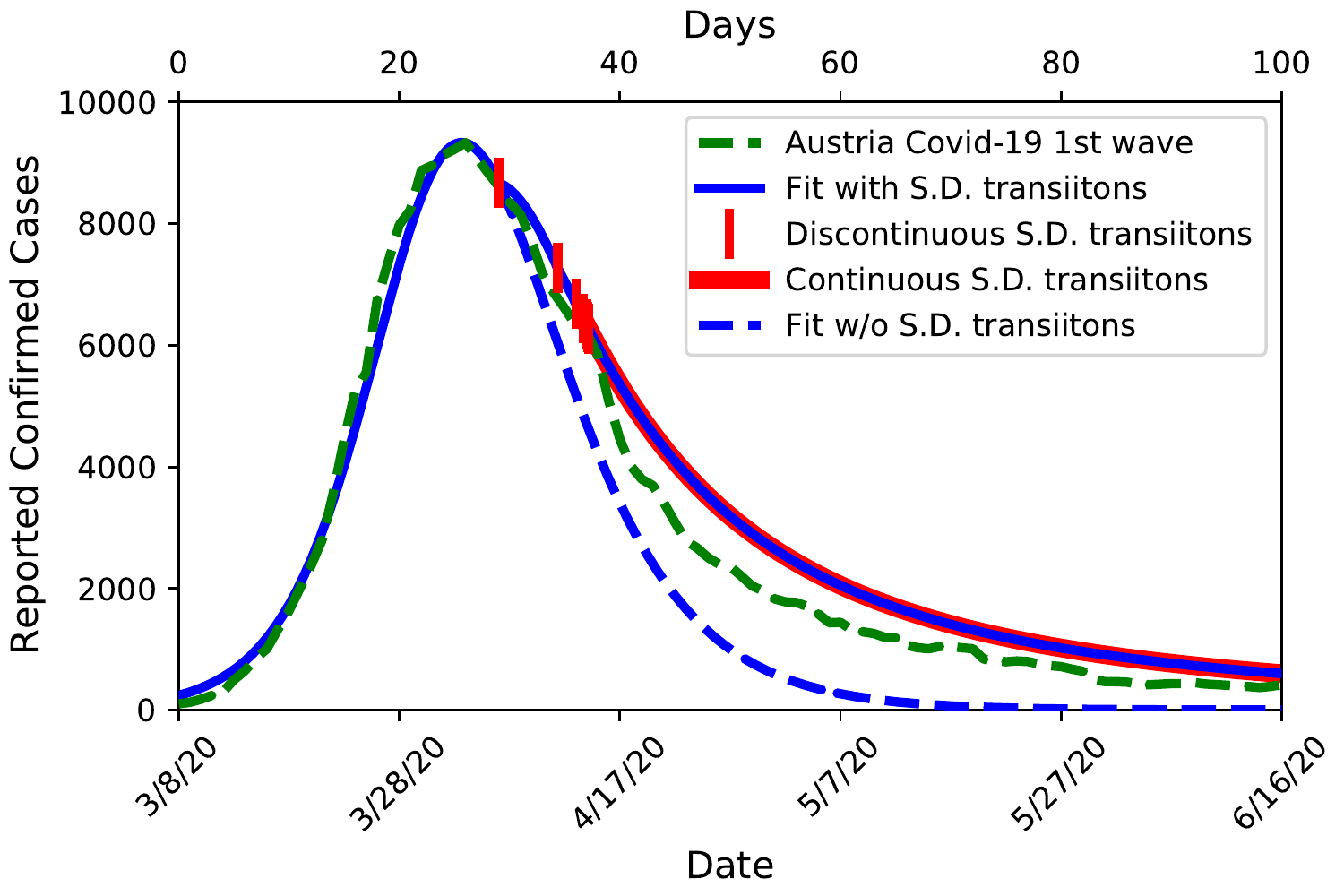}\includegraphics[scale=0.5]{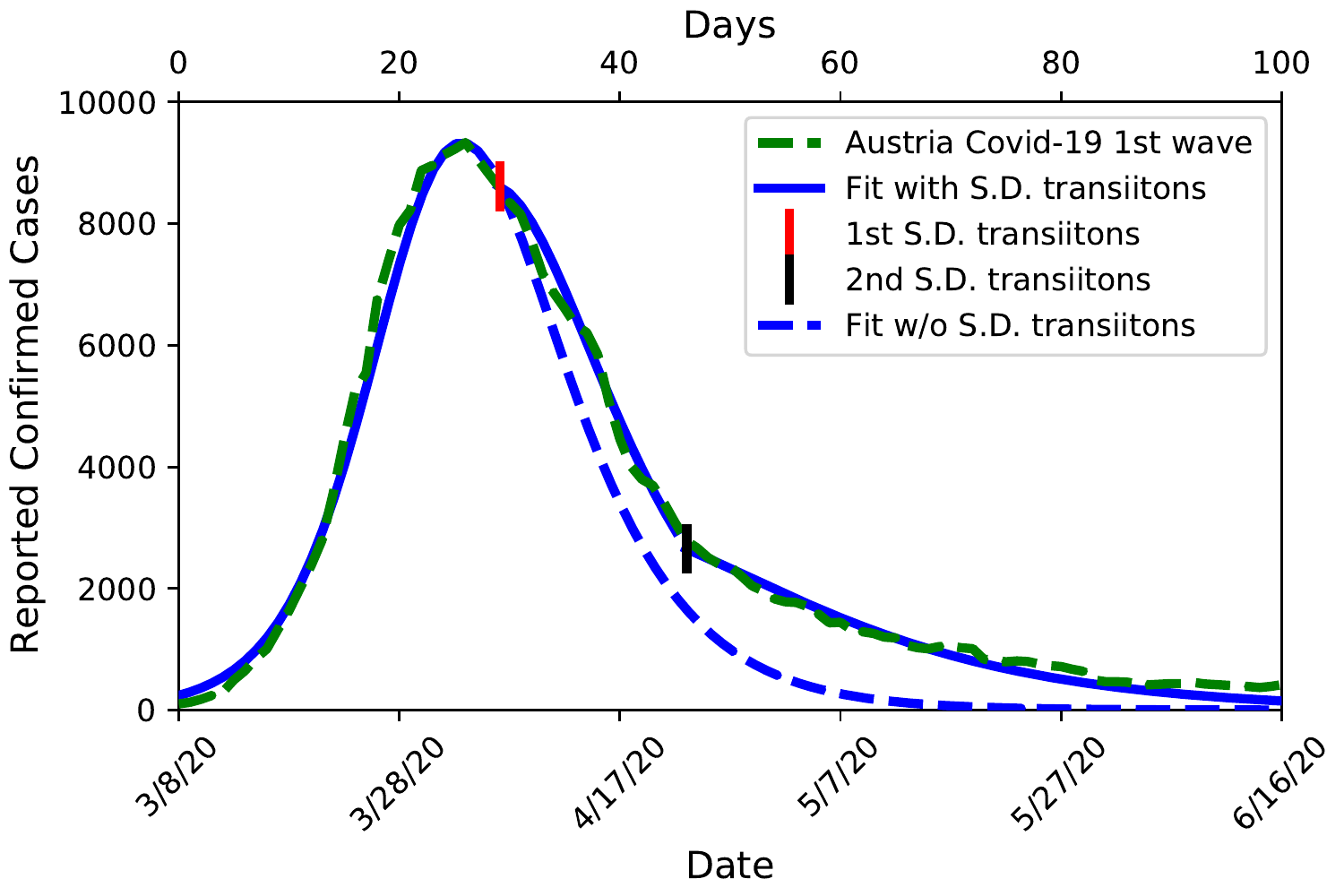}\caption{Alternative fits of Austria COVID-19 first wave A) fit with a different date for the first transition. There is a significant deviation between
reported and calculated confirmed cases. This demonstrates that the fit
techniques are sensitive to the choice of the first transition. This
sensitivity provides a hope that the fit may reveal something about
the real parameters
of the population or economy of the state. B) fit with two different distance
transitions. By adjusting utility function weights $E^{(1)}, E^{(2)}$ separately
for every transition a better fit can be achieved. In the case of Austria,
only two transitions are required. This method is less sensitive to
the choice of the first transition.}
\label{fig:4}
\end{figure}
\begin{figure}
\includegraphics[scale=0.5]{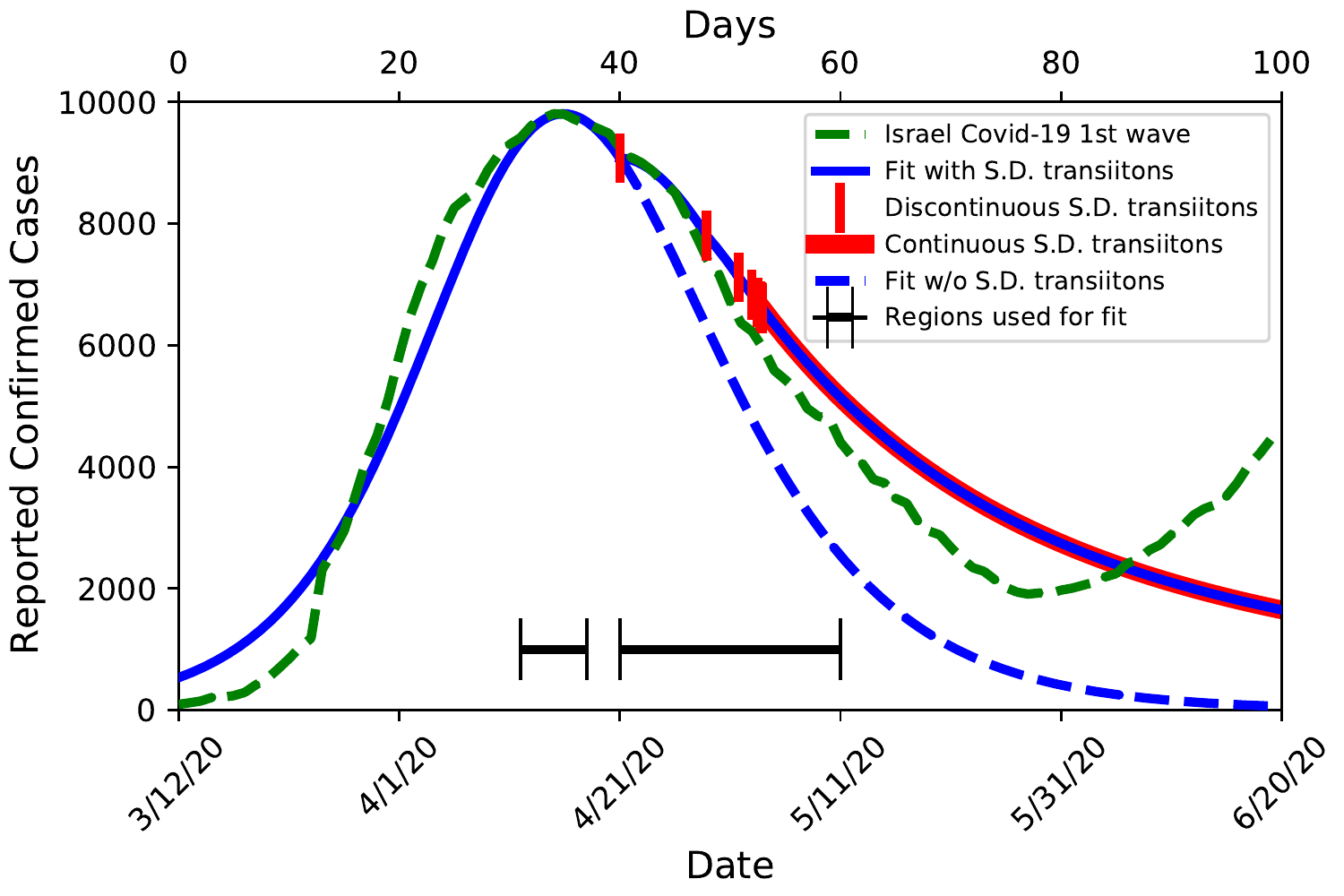}\includegraphics[scale=0.5]{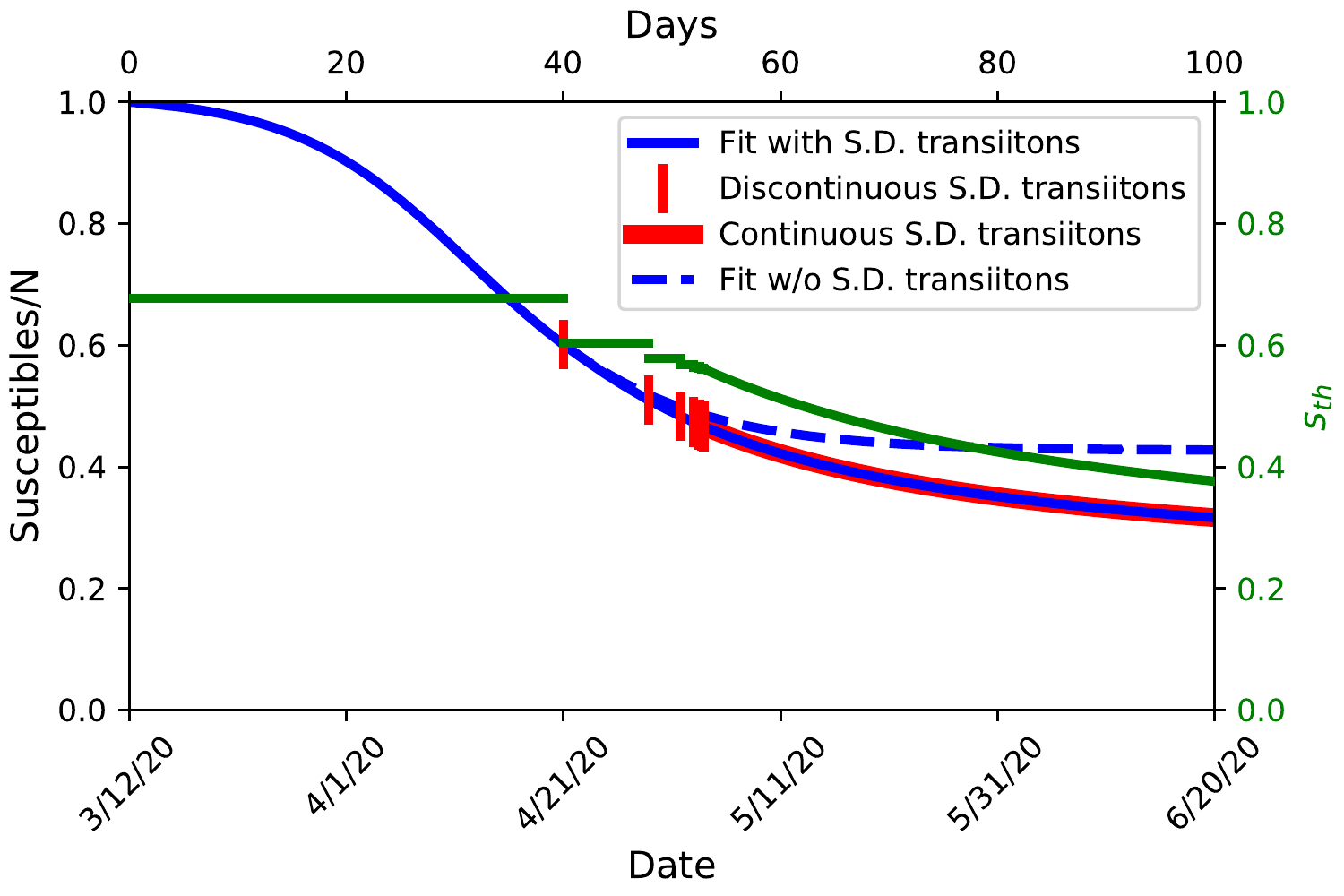}
\includegraphics[scale=0.5]{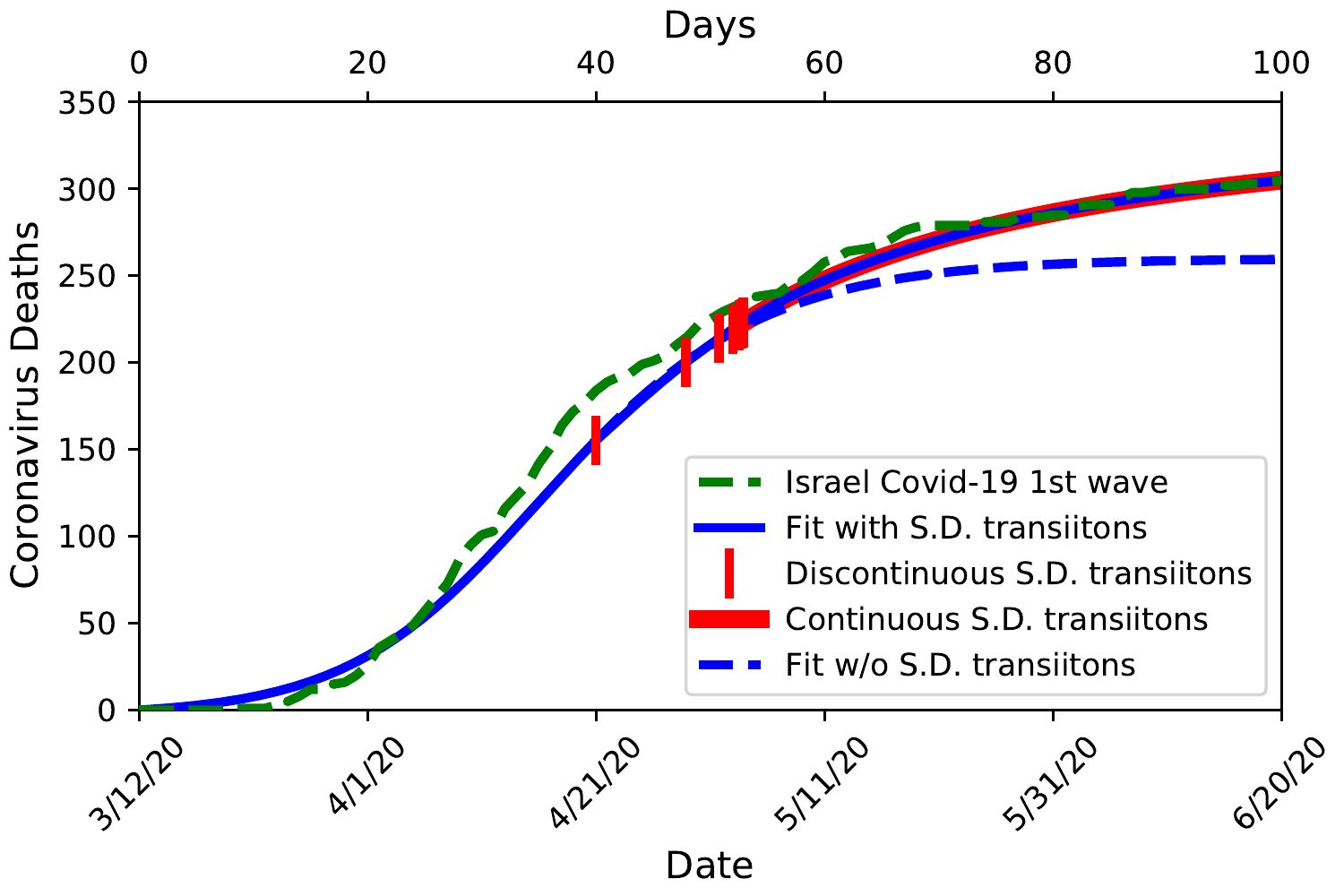}\includegraphics[scale=0.5]{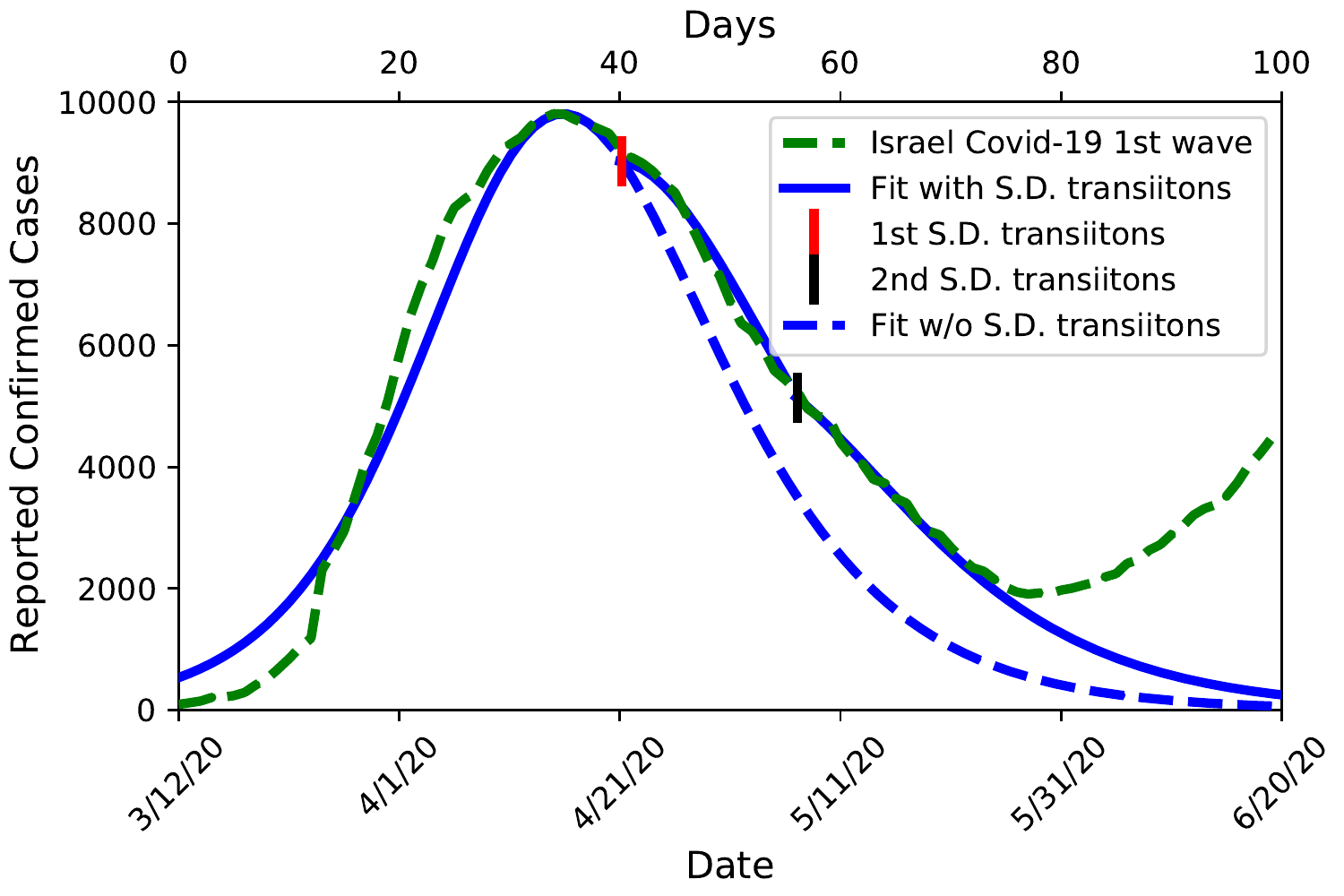}\caption{Fit of Israel's COVID-19 first wave. The results are similar to the case
of Austria. The fit is valid until the beginning of the second wave, at about
 the 70th day of the first one A) Confirmed cases. A significant deviation
exists between reported and calculated confirmed cases even before the
start of the second wave. B) Susceptible people and $s_{th}$. The social distancing
parameter $s_{th}$remains a bit higher in Israel relative to Austria
or Germany. C) Coronavirus deaths. The time delay between reported
and calculated cases is smaller than in the case of Austria. It can be
explained either by late or early reports of coronavirus tests or
reported deaths in Israel or Austria respectively. D) Alternative
fit with two transitions.}
\label{fig:5}
\end{figure}
\begin{figure}
\includegraphics[scale=0.5]{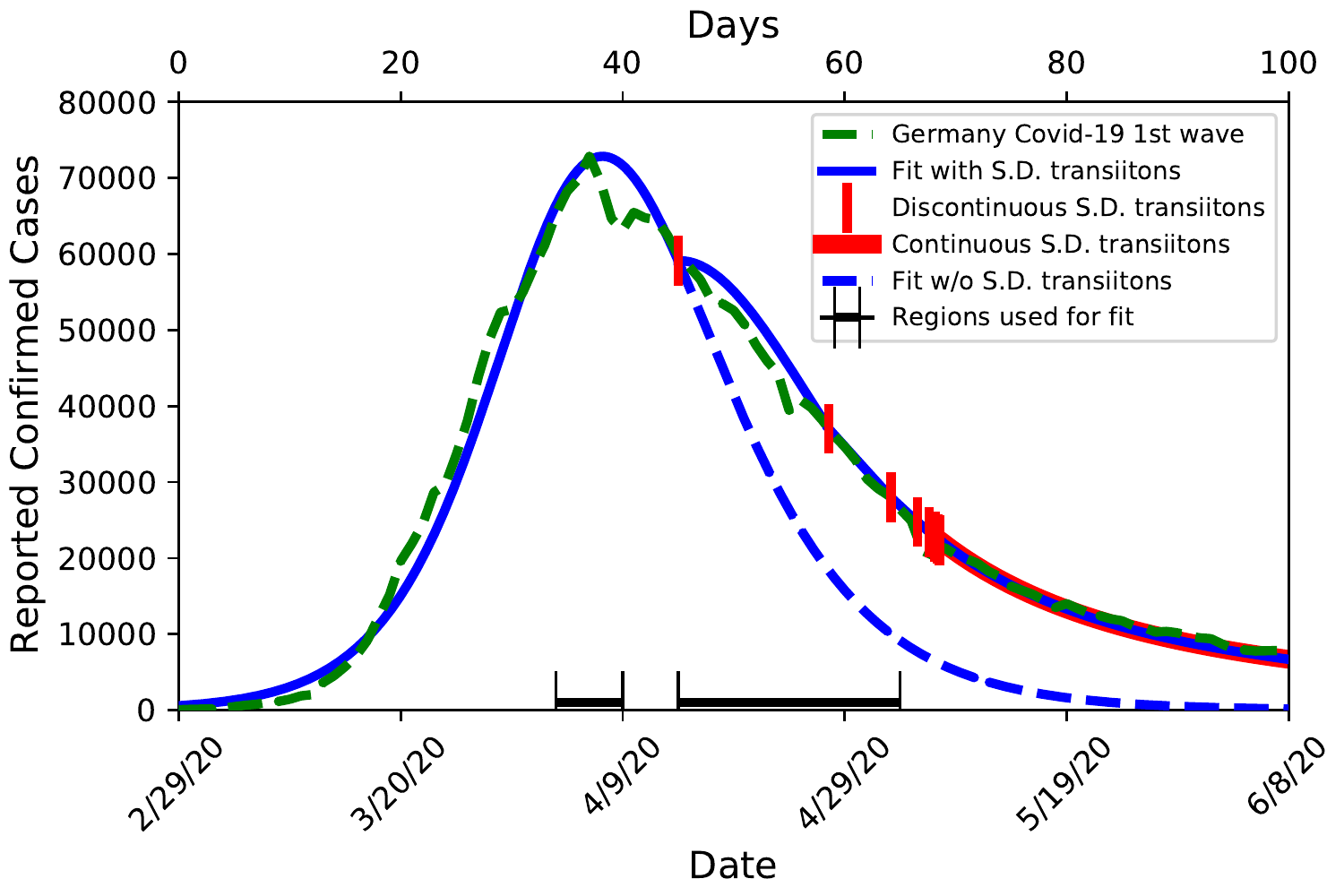}\includegraphics[scale=0.5]{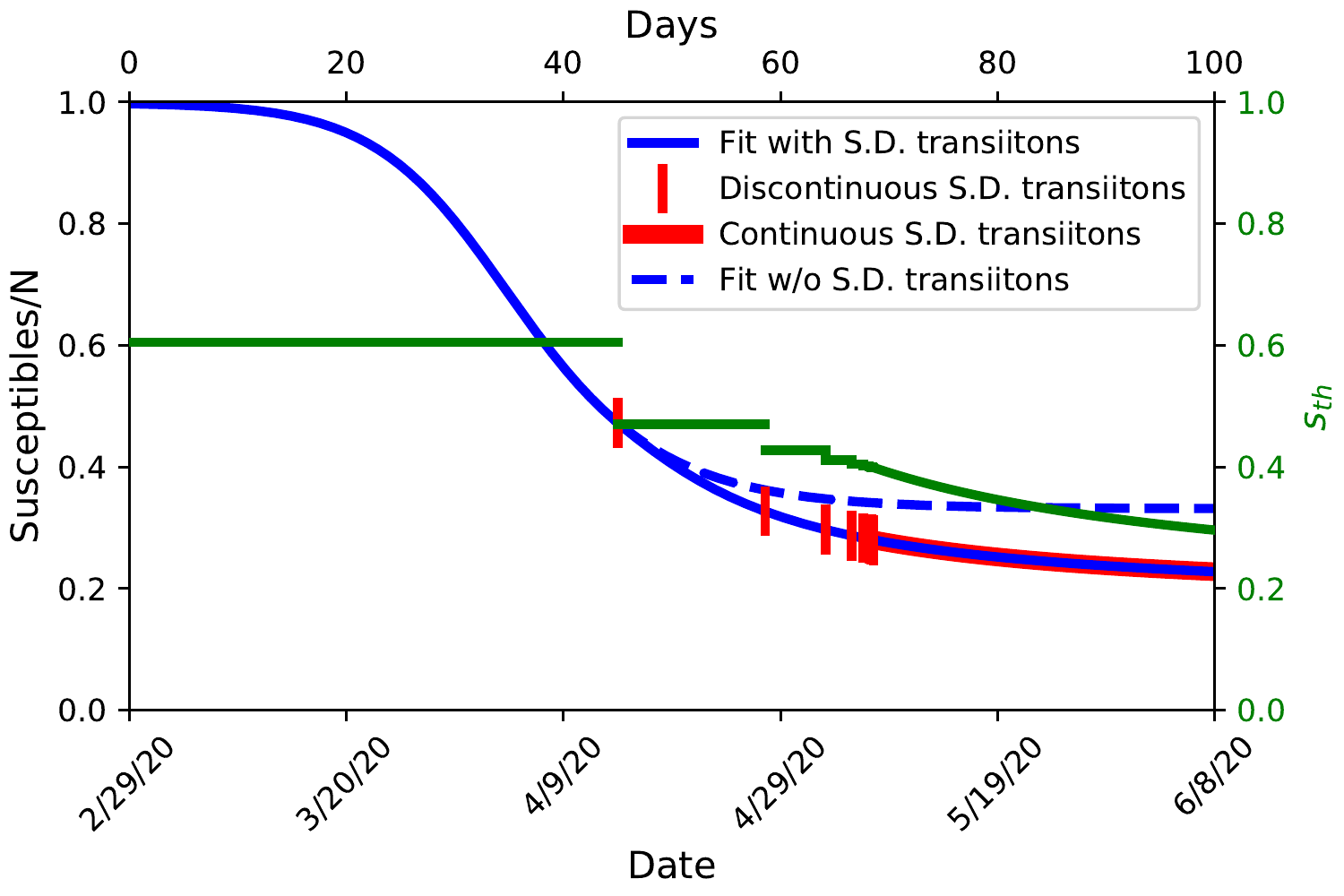}
\includegraphics[scale=0.5]{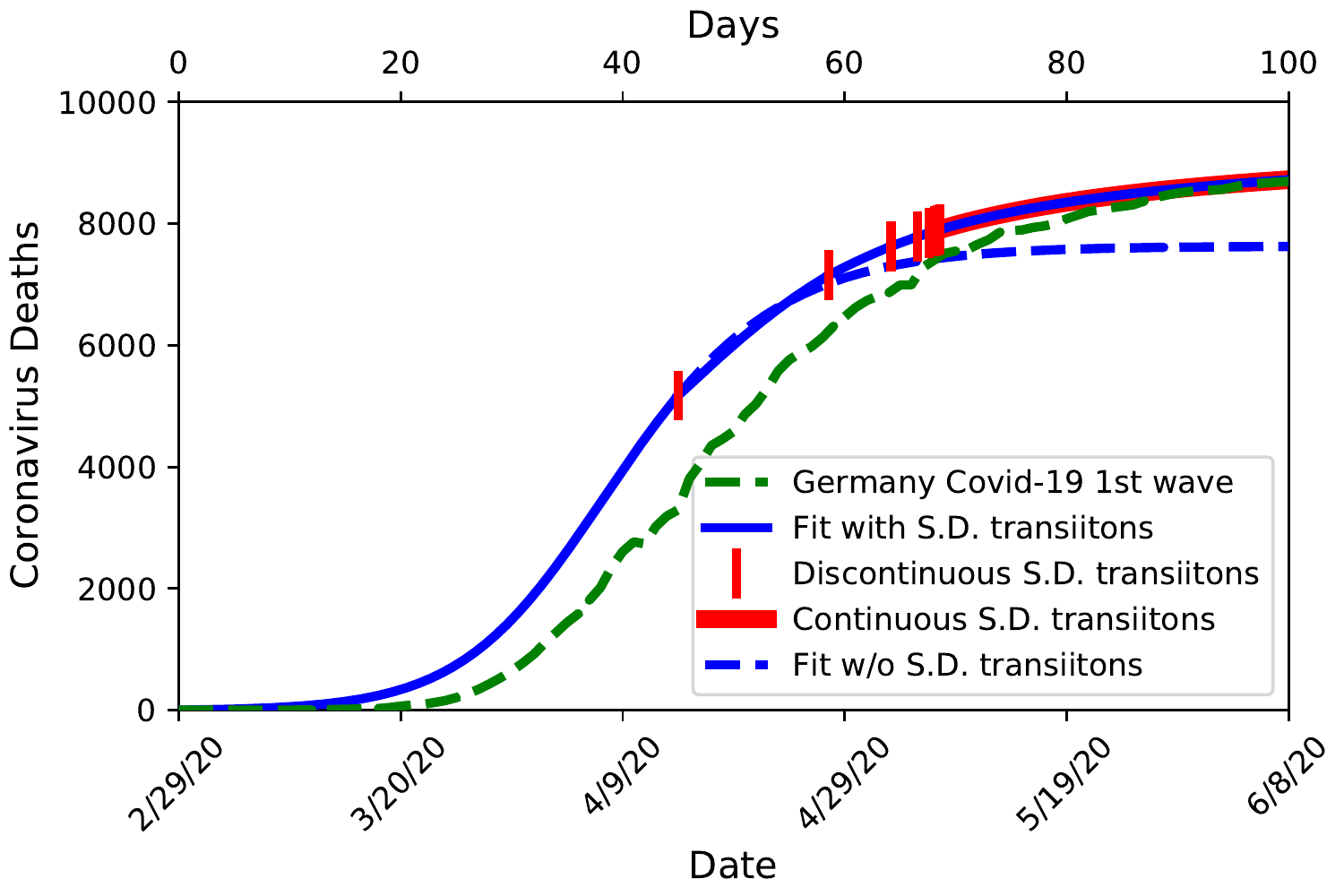}\includegraphics[scale=0.5]{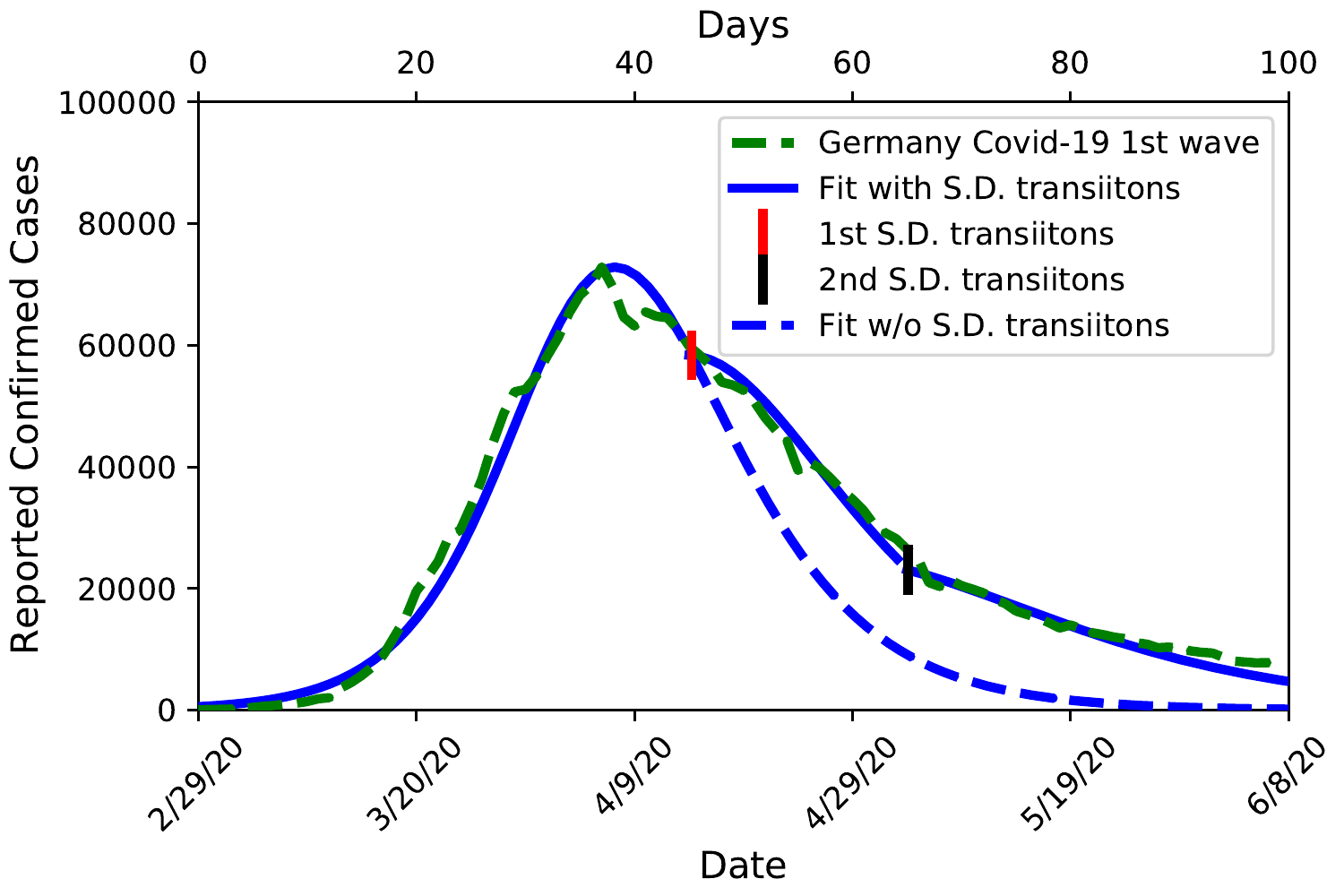}\caption{Fit of Germany's COVID-19 first wave. The results are very similar to
the case of Austria. A) Confirmed cases B) Susceptible people and $s_{th}$
C) Coronavirus deaths. There is a time delay between calculated and
reported deaths, like in the case of Austria. D) Alternative fit with
two transitions.}
\label{fig:6}
\end{figure}
\twocolumngrid
\end{widetext}

A small effective population size during the first wave is another
surprising result of the fit. An explanation of low $N$ may be that the initial
lockdown separated the population in disconnected
domains\cite{Callaway2000,Warren2002} and the wave of the epidemy occurred in a limited number of domains. The other possible
explanation is that a significant part of the population is immune
to COVID-19\cite{Mateus2020}. Finally, the SIR approach may be an oversimplified presentation of reality.

The small effective population size during the first wave may indicate a danger
of an abrupt transition to a bigger population size when $s_{th}$ reduces
below some critical value. It may result in a significant second wave
of the epidemy. It depends on the network structure of the population.

SIR dynamics can be mapped on
percolation in
an Erdos-Renyi network\cite{Grassberger1983}. $FES$ corresponds to a giant
component in percolation models. The critical value of the basic reproduction number for a percolation
transition was reported for some interaction
networks\cite{Sander2002,Warren2002,Newman2003a,Newman2002}, though
scale-free networks lack it\cite{Callaway2000}. Future work may extend
these tools to calculate the derivatives of $FES$ (\ref{eq:9}) to the general derivatives of
 giant components in percolation models.

Network theory may help to set the validity boundaries of this work. For
instance, the predictions of this work may become invalid
if the parameter of social distancing is a function $f(R_{0})$ such
that $FES$ as a function of $f$ ceases to be non-linear. An example of
such a function is the transmission
$T=1-\exp(-1/s_{th})$ in a scale-free network\cite{Newman2002}.

There are alternative explanations to discontinuities in observed
COVID-19 data, for instance, government
regulations\cite{Lopez2020} or continuous exogenous
phenomena\cite{Earn2000}. Relative weights of government regulations
and individual decision making regarding social distancing can not be
addressed in the framework of this work. Nevertheless, testing the basic decision-making
model against real data is important. The evolution of decision making
could converge to some basic model which is different from complex
modern reality.

To conclude, this
work predicts observable discontinuous transitions of social distancing and provides tools for quantitative
analysis of pandemic waves. The dynamics include the Zeno-like effect of infinite transitions during a finite time.
Multiple transitions may be interesting for theories of
crisis formation - the probability that something goes wrong increases with
the number of decision points. The developed tools
contributing to social epidemiology, like the SIR model, may be mapped to network
percolation theory and to the spread of
non-contagious but going-viral phenomena\cite{Shiller2017,Bauch2013a}.

\begin{widetext}
\onecolumngrid
\begin{table}[h]
\begin{tabular}{|l|c|c|c|c|c|c|c|c|c|c|c|c|c|c|c|c|c|c|c|}
\hline 
 & $s_{0}$ & $i_{0}$ & $\gamma$ & $s_{th}$ & $\beta$ & $s_{th}^{(1)}$ &
                                                                      $\beta^{1}$ & $s_{th}^{(2)}$ & $\beta^{2}$ & $A$ & $N$ & $A'$ & $M\;(IFR)$  & $D_{1}$ & $D_{2}$ & $E^{(1)}$ & $E^{(2)}$\tabularnewline
\hline 
\hline 
Austria, Figs. \ref{fig:22}, \ref{fig:223c},\ref{fig:3} & 0.99 & 242 & 0.26 & 0.56 & 0.46 & 0.45 &
                                                                     0.46
  &  &  & 1.26e-05 & 8.2e5$/k$ & 10$/k$ & $k\times 10^{-3}$ & 32 &  & 1.31 & 5.81 \tabularnewline
\hline 
Austria, Fig. \ref{fig:3} A & 0.99 & 242 & 0.26 & 0.56 & 0.46 & 0.48 &
                                                                       0.46 &  &  & 1.26e-05 & 8.2e5$/k$ & 10$/k$ & $k\times 10^{-3}$  & 29 &  & 1.84 & 6.30 \tabularnewline
\hline 
Austria, Fig. \ref{fig:3} B & 0.99 & 242 & 0.26 & 0.56 & 0.46 & 0.48 & 0.46 & 0.32 & 0.46 & 1.26e-05 & 8.2e5$/k$ & 10$/k$ & $k\times 10^{-3}$  & 29 & 46 & 1.84 & 6.30 \tabularnewline
\hline 
Germany, Fig. \ref{fig:6} A,B,C & 0.99 & 578 & 0.26 & 0.60 & 0.43 &  &
  &  &  & 1.24e-6 & 1.1e7$/k$ & 14$/k$ & $k\times 10^{-3}$   & 45 &  & 1.15 & 5.93  \tabularnewline
\hline 
Germany, Fig. \ref{fig:6} D & 0.99 & 578 & 0.26 & 0.60 & 0.43 & 0.47 & 0.43 & 0.33 & 0.52 & 1.24e-6 & 1.1e7$/k$ & 14$/k$ & $k\times 10^{-3}$  & 45 & 65 & 1.15 & 5.93 \tabularnewline
\hline 
Israel, Fig. \ref{fig:5} A,B,C & 1.0 & 535 & 0.26 & 0.68 & 0.38 & 0.60 & 0.38 &  &  & 6.34e-06 & 4.5e5$/k$ & 3$/k$ & $k\times 10^{-3}$  & 40 &  & 1.33 & 4.92 \tabularnewline
\hline 
Israel, Fig. \ref{fig:5} D & 1.0 & 535 & 0.26 & 0.68 & 0.38 & 0.60 & 0.38 & 0.50 & 0.51 & 6.34e-06 & 4.5e5$/k$ & 3$/k$ & $k\times 10^{-3}$  & 40 & 56 & 1.33 & 4.92 \tabularnewline
\hline
\end{tabular}\caption{Parameters of SIRIT model that fit the first COVID-19 wave of Austria,
Germany, and Israel. The parameters may be used to reproduce the figures
of the article. For definition of parameters see
eqs. (\ref{eq:mainsirit0}), (\ref{eq:mainsirit1}) and (\ref{eq:mainsirit2}).
Besides, $D_{1}$and $D_{2}$ are the days of the first and the
second (if required) transitions. $E^{(1)},E^{(2)}$
are the weights of the utility function which were fitted for the first
transition. An interesting result is a small effective
population size $N$, less than $1/10$ of the state population. Greater
values of mortality rate $M$ (factor $k$) predict even lower values of population
size $N$ (increase in $M$ causes a proportional reduction in effective
population size $N$ and
$A'$ ratio between the real and reported number of
infected). Mortality rate $M>0.3\%$ ($k>3$) causes non-physical $A'<1$ in the case of
Israel. The second
transition in the case of Germany and Israel brings a change in $\beta$.
The results of all three countries are quite similar except $A'$ -
ratio between the real and reported number of infected people.}
\label{tab:1}
\end{table}
\twocolumngrid
\end{widetext}

\appendix

\section{Analytic solution of (\ref{eq:mainsirit1})}
\label{sec:analyt-solut-refeq:m}

The first two equations of (\ref{eq:mainsirit0}) may be rewritten
as:
\begin{equation}
\frac{1}{\beta}\frac{\partial z}{\partial x}z=-\beta A\left(s_{th}+\frac{1}{\beta}z\right)\exp\left(x\right),\label{eq:egdgd}
\end{equation}
using transformation $\frac{\partial\log I}{\partial t}=z,\frac{\partial z}{\partial t}=\frac{\partial z}{\partial\log I}\frac{\partial\log I}{\partial t}=\frac{\partial z}{\partial\log I}z,x=\log I$.
Integration of (\ref{eq:egdgd}) results in:
\begin{equation}
i=i_{0}+\left[s_{0}-s+s_{th}\log\left[\frac{s}{s_{0}}\right]\right],\label{eq:rttrtr}
\end{equation}
where $(s_{0},i_{0})$ are initial values of $s$ and $i$.

The time between current value $s$ and $s_{tr}$ is: 
\begin{equation}
\int_{s}^{s_{tr}}\frac{ds}{\frac{ds}{dt}}=\int_{s}^{s_{tr}}\frac{ds}{-\beta \left[i_{0}+\left[s_{0}-s+s_{th}\log\left[\frac{s}{s_{0}}\right]\right]\right]s},\label{eq:ertre}
\end{equation}
and remain finite while $s_{tr}>s_{min}$. 

To derive $s$ in the form of the Lambert $W$ function (defined
as $W(x)$, where $W\exp W=x$) one
should rewrite (\ref{eq:rttrtr}) as:
\begin{eqnarray}
\label{eq:2}
  -\frac{s}{s_{th}}\exp\left[-\frac{s}{s_{th}}\right] & =-\frac{s_{0}}{s_{th}}\exp\left[\frac{\left(i-i_{0}\right)-s_{0}}{s_{th}}\right].
\end{eqnarray}
The expressions (\ref{eq:2}) and (\ref{eq:rttrtr}) provide
connections between $s$ and $i$ along the population trajectory in $(s,i)$
space that initiates at $(s_{0},i_{0})$.  Lambert W  should be used
carefully because it is a multi-valued function.

Let us define by $(s_{t},i_{t})$ the values of the susceptible ratio $s$
and the number of infected $i$ along the population trajectory in $(s,i)$
space that initiates at $(s_{0},i_{0})$. Then let us calculate the
coefficients $F^{(1)},F^{(2)},F^{(3)}$~(\ref{eq:8}) and show that along trajectory
$(s_{t},i_{t})$ they are polynomials of $\log\left[s_{t}\right]$.
Derivatives of $\log\left[s_{\min}\right]$:
\begin{equation}
\log\left[s_{min}\right]=\log\left[s_{th}\right]\text{+\ensuremath{\log}\ensuremath{\left[W\left(f(s_{th})\right)\right]},}\label{eq:5ytrry}
\end{equation}
due to $s_{th}$ include derivatives of $\log\left[s_{th}\right]$
and derivatives of $\ensuremath{\log}\left[W\left(f(s_{th})\right)\right]$,
where:
\begin{equation}
f(s_{th})=-\frac{\text{s}_{0}}{s_{th}}\exp\left[-\frac{s_{0}+i_{0}}{s_{th}}\right].\label{eq:rege}
\end{equation}

Let us notice that $s_{th}$, $f(s_{th})$ and $W\left(f(s_{th})\right)$
are constant along trajectory $(s_{t},i_{t})$ until the
value of $s_{th}$changes
by a transition. The values $f(s_{th})$ and $W\left(f(s_{th})\right)$:
\begin{eqnarray}
f & =-\frac{\text{s}_{0}}{s_{th}}\exp\left[-\frac{s_{0}}{s_{th}}-\frac{i_{0}}{s_{th}}\right],\\
W(f) & =-\frac{s_{\min}}{s_{th}},
\label{eq:1}\end{eqnarray}
are constant until a transition takes place, because $s_{min}$ (\ref{eq:5ytrry}) is constant along the trajectory
unless $s_{th}$changes its value. Besides:
\begin{equation}
-\frac{s_{0}}{s_{th}}\exp\left[-\frac{i_{0}+s_{0}}{s_{th}}\right]=\left(-\frac{\text{s}_{t}}{s_{th}}\exp\left[-\frac{s_{t}}{s_{th}}-\frac{i_{t}}{s_{th}}\right]\right)=const\label{eq:etr}
\end{equation}
because any point $(s_{t},i_{t})$ can serve as an initial value $(s_{0},i_{0})$
for the continuation of the trajectory.

Derivatives of $\log\left[s_{th}\right]$ depend only on $s_{th}$
and, so, are constant until $s_{th}$ changes. 

Derivatives of $\ensuremath{\log}\left[W\left(f(s_{th})\right)\right]$
are:
\begin{eqnarray}
\frac{d\ensuremath{\log}W\left(f\right)}{ds_{th}} &=&\frac{d\ensuremath{\log}W\left(f\right)}{df}\frac{df}{ds_{th}}\nonumber\\
\frac{d\ensuremath{^{2}\log}W\left(f\right)}{ds_{th}^{2}} &=&\frac{d\ensuremath{^{2}\log}W\left(f\right)}{df^{2}}\left(\frac{df}{ds_{th}}\right)^{2}+\frac{d\ensuremath{\log}W\left(f\right)}{df}\frac{d^{2}f}{ds_{th}^{2}}\nonumber\\
\frac{d^{3}\ensuremath{\log}W\left(f\right)}{ds_{th}^{3}}
                                                  &=&\frac{d\ensuremath{^{3}\log}W\left(f\right)}{df^{3}}\left(\frac{df}{ds_{th}}\right)^{3}+\nonumber\\
  &&3\frac{d\ensuremath{^{2}\log}W\left(f\right)}{df^{2}}\frac{df}{ds_{th}}\frac{d^{2}f}{ds_{th}^{2}}+ \frac{d\ensuremath{\log}W\left(f\right)}{df}\frac{d^{3}f}{ds_{th}^{3}}\nonumber\\
\end{eqnarray}
Derivatives of $\ensuremath{\log}W\left(f\right)$ due to $f$ are
invariant along $(s_{t},i_{t})$ trajectory, see Appendix \ref{sec:deriv-lamb-w}.

Derivatives of $f$ due to $s_{th}$are polynomials of $\log\left[s_{t}\right]$.
Consider the first derivative of (\ref{eq:rege}) taking into account
(\ref{eq:etr}):
\begin{eqnarray}
  \frac{df}{ds_{th}}=gf,
\end{eqnarray}
where:
\begin{equation}
g=\left(-\frac{1}{s_{th}}\right)+\left(\frac{s_{t}+i_{t}}{s_{th}^{2}}\right).\label{eq:rwert}
\end{equation}
Expression (\ref{eq:rttrtr}) can be rewritten in the form:
\begin{equation}
i_{t}+s_{t}=i_{0}+s_{0}+s_{th}\log\left[\frac{s_{t}}{s_{0}}\right]\label{eq:trte}
\end{equation}
for trajectory $(s_{t},i_{t})$. Plugging (\ref{eq:trte}) in (\ref{eq:rwert})
results in:
\begin{equation}
g=\left[\left(-\frac{1}{s_{th}}\right)+\left(\frac{i_{0}+s_{0}+s_{th}\log\left[\frac{s_{t}}{s_{0}}\right]}{s_{th}^{2}}\right)\right].
\end{equation}
Thus $g$ and its derivatives due to $s_{th}$ are linear functions
of $\log\left[s_{t}\right]$. The second and the third derivatives
then:
\begin{equation}
\frac{d^{2}f}{ds_{th}^{2}}=fg^{2}+f\frac{dg}{ds_{th}}
\end{equation}
\begin{eqnarray}
\frac{d^{3}f}{ds_{th}^{3}} & =fg^{3}+3fg\frac{dg}{ds_{th}}+f\frac{d^{2}g}{ds_{th}^{2}}
\end{eqnarray}
are the quadratic and cubic polynomials of $\log\left[s_{t}\right]$.

\section{Derivatives of the Lambert $W$ function}
\label{sec:deriv-lamb-w}
All derivatives of $W$ due to $f$ are constant along any trajectory in
$(s,i)$ space: 

\begin{eqnarray}
W\exp\left[W\right] & =f,
\end{eqnarray}
\begin{equation}
\frac{d\log W}{df}=\frac{1}{f(1+W)},
\end{equation}
\begin{eqnarray}
\frac{d^{2}\log W}{df^{2}} & =-\frac{1}{f}\frac{d\log W}{df}-fW\left[\frac{d\log W}{df}\right]^{3},
\end{eqnarray}
\begin{eqnarray}
\frac{d^{3}\log W}{df^{3}} && =\frac{1}{f^{2}}\frac{d\log
                             W}{df}-\frac{1}{f}\frac{d^{2}\log
                             W}{df^{2}}-W\left[\frac{d\log
                              W}{df}\right]^{3}-\nonumber\\
  &&fW\left[\frac{d\log W}{df}\right]^{4}-3fW\left[\frac{d\log W}{df}\right]^{2}\frac{d^{2}\log W}{df^{2}}.\nonumber\\
\end{eqnarray}
The final expressions are invariant until $s_{th}$ changes because
they depend on $f$ and W only, see (\ref{eq:1}).

\section{Expressions for $F^{(1)}$, $F^{(2)}$ and  $F^{(3)}$}
\label{sec:final-expressions-b}
The first:
\begin{equation}
F^{(1)}=-\frac{i_{0}-s_{\min}+s_{\text{th}}\log\left(\frac{s}{s_{0}}\right)+s_{0}}{s_{\text{th}}\left(s_{\text{th}}-s_{\min}\right)}
\end{equation}

The second:
\begin{eqnarray}
  &&F^{(2)}=-\frac{1}{2s_{\text{th}}^{2}\left(s_{\text{th}}-s_{\min}\right){}^{3}}\times\\
  &&\left(i_{0}-s_{\min}+s_{\text{th}}\log\left(\frac{s}{s_{0}}\right)+s_{0}\right)\times\nonumber\\
&&\left(s_{\min}\left(i_{0}-s_{\min}+s_{0}\right)+  2s_{\min}s_{\text{th}}+s_{\min}s_{\text{th}}\log\left(\frac{s}{s_{0}}\right)-2s_{\text{th}}^{2}\right)\nonumber
\end{eqnarray}

and
\begin{widetext}
\begin{eqnarray}
F^{(3)} &
    =&-\frac{1}{6s_{\text{th}}^{3}\left(s_{\text{th}}-s_{\min}\right){}^{5}}\Bigg
       [\left(i_{0}-s_{\min}+s_{\text{th}}\log\left(\frac{s}{s_{0}}\right)+s_{0}\right)\left(-3s_{\min}s_{\text{th}}^{2}\left(3i_{0}-5s_{\min}+3s_{0}\right)\right)+\nonumber\\
&& s_{\min}s_{\text{th}}\left(i_{0}-s_{\min}+s_{0}\right)\left(i_{0}+8s_{\min}+s_{0}\right)+s_{\min}s_{\text{th}}\log\left(\frac{s}{s_{0}}\right)\times\\\nonumber
&& \left(s_{\text{th}}\left(2i_{0}+7s_{\min}+2s_{0}\right)+4s_{\min}\left(i_{0}-s_{\min}+s_{0}\right)+s_{\text{th}}\log\left(\frac{s}{s_{0}}\right)\left(2s_{\min}+s_{\text{th}}\right)-9s_{\text{th}}^{2}\right)+\\\nonumber
&&
     2s_{\min}^{2}\left(i_{0}-s_{\min}+s_{0}\right){}^{2}-12s_{\min}s_{\text{th}}^{3}+6s_{\text{th}}^{4}\Bigg ]
\end{eqnarray}
\end{widetext}
where $s_{min}$ is defined by (\ref{eq:5ytrry}).

Expressions developed
in appendices use the parameters of the SIR model
eqs. (\ref{eq:mainsirit0}). Transformation $i\rightarrow AI$ converts
previous equations to the parameters of (\ref{eq:mainsirit2}).
\bibliographystyle{apsrev}
% \bibliography{/mnt/c/Users/sasha/GoogleDrive/Articles/corona,/mnt/c/Users/sasha/GoogleDrive/Articles/SirCost,/mnt/c/Users/sasha/GoogleDrive/Articles/Regula2010refs,/mnt/c/Users/sasha/GoogleDrive/Articles/SanderRefs,/mnt/c/Users/sasha/GoogleDrive/Articles/corona_mortality,/mnt/c/Users/sasha/GoogleDrive/Articles/phasetransocial,/mnt/c/Users/sasha/GoogleDrive/Articles/coronaSocialDist,/mnt/c/Users/sasha/GoogleDrive/Articles/coronaSocDist,/mnt/c/Users/sasha/GoogleDrive/Articles/coronaPercolation}

\end{document}